\documentclass[letter,apj]{emulateapj-rtx4}
\usepackage{natbib}
\usepackage{amssymb,amsmath}
\usepackage{graphicx}

\def\figheight{8.0cm}
\def\ud{\,\mathrm{d}}

\begin{document}

\title{Results From Core-Collapse Simulations with
  Multi-Dimensional, Multi-Angle Neutrino Transport}
\shorttitle{Core-Collapse Simulations}
\author{\bigskip\bigskip Timothy D. Brandt\altaffilmark{1}, 
Adam Burrows\altaffilmark{1}, Christian D. Ott\altaffilmark{2}, and
  Eli Livne\altaffilmark{3}}

\altaffiltext{1}{Department of Astrophysical Sciences, Peyton Hall,
  Princeton University, Princeton, NJ 08544, USA \\
  tbrandt@astro.princeton.edu, burrows@astro.princeton.edu}
\altaffiltext{2}{Theoretical Astrophysics, Mail Code 350-17,
  California Institute of Technology, Pasadena, CA  \\ cott@tapir.caltech.edu}
\altaffiltext{3}{Racah Institute of Physics, Hebrew University,
  Jerusalem, Israel  \\ livne@phys.huji.ac.il}

\begin{abstract}
We present new results from the only 2D multi-group, multi-angle
calculations of core-collapse supernova evolution.  The first set of
results from these calculations was published in Ott et al. (2008).
We have followed a nonrotating and a rapidly rotating 20-$M_\odot$
model for $\sim$400 ms after bounce.  We show that the radiation
fields vary much less with angle than the matter quantities in the
region of net neutrino heating.  This obtains because most neutrinos
are emitted from inner radiative regions and because the specific
intensity is an integral over sources from many angles at depth.  The
latter effect can only be captured by multi-angle transport.  We then
compute the phase relationship between dipolar oscillations in the
shock radius and in matter and radiation quantities throughout the
postshock region.  We demonstrate a connection between variations in
neutrino flux and the hydrodynamical shock oscillations, and use a
variant of the Rayleigh test to estimate the detectability of these
neutrino fluctuations in IceCube and Super-K.  Neglecting flavor
oscillations, fluctuations in our nonrotating model would be
detectable to $\sim$10 kpc in IceCube, and a detailed power spectrum
could be measured out to $\sim$5 kpc.  These distances are
considerably lower in our rapidly rotating model or with significant
flavor oscillations.  Finally, we measure the impact of rapid rotation
on detectable neutrino signals.  Our rapidly rotating model has
strong, species-dependent asymmetries in both its peak neutrino flux
and its light curves.  The peak flux and decline rate show
pole-equator ratios of up to $\sim$3 and $\sim$2, respectively.
\end{abstract}


\section{Introduction}

The basic context of a core-collapse supernova is well-established:
a collapsing, degenerate core releases an enormous amount of energy,
$\sim$1\% of which couples to and unbinds the stellar envelope.
Despite decades of research, the mechanism of this coupling remains
obscure.  The collapsing core rebounds at nuclear densities and
launches a bounce shock, but detailed simulations in spherical
symmetry show that the shock wave stalls at 100-200 km and fails to
explode the star \citep{RamppJanka00, LiebendorferEtAl01,
  LiebendorferEtAl05, ThompsonEtAl03}.  The best 2D simulations
confirm this result; much of the shock's energy is lost to nuclear
dissociation and escaping neutrinos, and the shock stalls
\citep{Bethe90, JankaEtAl07}.

The neutrino mechanism, first proposed by \cite{ColgateWhite66} in its
prompt form, posits a burst of neutrino emission to energize the
shock.  In its delayed form \citep{BetheWilson85}, an imbalance
between neutrino absorption and emission behind the shock deposits the
additional required energy over several hundred milliseconds.  The
rate of energy deposition in this ``gain region'' depends on the
relationship between the neutrino flux, which declines as $\sim
r^{-2}$, and the cooling rate $\overline{\kappa}acT^4 \sim T^6$, which
generally falls off much more quickly. 

Early 2D simulations successfully exploded supernova progenitors by
the delayed neutrino mechanism \citep{HerantEtAl94, BurrowsEtAl95,
JankaMuller96, FryerHeger00, FryerWarren02, FryerWarren04}, though
more recent calculations with better neutrino transport generally fail
to obtain explosions.  An exception is for $\sim$8--9 $M_\odot$ stars,
whose steep density gradients reduce the ram pressure of infalling
matter.  The post-shock pressure needed to drive an explosion abruptly
drops, and a sub-energetic wind-driven supernova follows
\citep{KitauraEtAl06, BurrowsEtAl07a}.  There are indications that
more generic 2D explosions may be obtained with a soft nuclear
equation of state, which results in a compact neutron star
\citep{BruennEtAl10, MarekJanka09}.  However, recent experiments
\citep{ShlomoEtAl06} and observations of massive neutron stars
\citep{DemorestEtAl10} favor a stiffer equation of state.
\citeauthor{BruennEtAl10}~also find the shock to stall at a much
larger radius than other groups, for reasons that are not yet clear.
See \cite{NordhausEtAl10} and references therein for a more thorough
discussion of the state of 2D simulations.

The delayed neutrino mechanism requires $\sim$10\% of the energy
emitted by the core in electron and anti-electron neutrinos over the
first few hundred milliseconds, a few $10^{51}$ erg, to be deposited
in the post-shock material.  The details of this energy deposition
depend both on the neutrino-matter coupling and on the hydrodynamics,
which determines how long individual particles are subjected to net
neutrino heating.  This latter point has been shown to be a strong
function of dimension, with \cite{MurphyBurrows08} and
\cite{NordhausEtAl10} demonstrating that explosions require $\sim$30\%
less neutrino heating in 2D than in 1D, and $\sim$15--25\% less in 3D
than in 2D.  Two-dimensional core-collapse simulations display
powerful, low-mode oscillations of matter behind the stalled shock as
a result of the Standing Accretion Shock Instability
\citep[SASI,][]{FoglizzoTagger00, BlondinEtAl03, FoglizzoEtAl07,
IwakamiEtAl08, YamasakiFoglizzo08, ScheckEtAl08}.  Recent 3D
calculations confirm the presence of these oscillations but find their
energy to be spread over a wide range of modes forbidden in
axisymmetry \citep{FryerYoung07, Fernandez10, NordhausEtAl10}.  The
coupling of neutrinos to this hydrodynamically unstable matter is
central to the supernova problem.

In this work, we present new, hitherto unpublished features of the
only 2D multi-group, multi-angle neutrino transport calculations of
core-collapse supernova evolution ever performed.  Results from these
simulations were first published by \cite{OttEtAl08}.  These authors
studied one nonrotating and one rapidly rotating model, and discussed
the overall behavior of the evolution and differences between models
evolved using full multi-angle transport and those using multi-group
flux-limited diffusion.  \citeauthor{OttEtAl08}~also studied the
angular character of the specific intensity, a quantity that may only
be calculated using multi-angle transport.  We extend these results,
characterizing the spatial variation and coupling of the matter and
radiation fields.  We also detail the spatial and temporal variations
in estimated neutrino signals observable from Earth.  We obtain
similar results to \cite{MarekEtAl09} and \cite{LundEtAl10}, but with
more detailed neutrino transport and a different statistical
technique.

\cite{OttEtAl08} demonstrated that multi-angle transport
self-consistently captures the transition from diffusion to
free-streaming, eliminating flux-limiter artifacts (though introducing
artifacts intrinsic to $S_n$ in the optically thin limit; see
\cite{Castor04}).  They also showed that the net neutrino heating
rates behind the stalled shock are higher in their nonrotating model
with multi-angle transport than with flux-limited diffusion.  Though
the difference ranges from $\sim$5--10\% at early times to
$\sim$20--30\% at later times, \citeauthor{OttEtAl08} do not find
their 2D multi-angle model to be significantly closer to explosion.

This paper complements \citeauthor{OttEtAl08}~by examining in detail
the coupling between matter and radiation and the relationship between
spatial and temporal fluctuations in hydrodynamic quantities and
neutrino spectral energy densities.  We find that the magnitude of
fluctuations is substantially lower in the radiation fields than in
the local matter in the critical region behind the stalled shock.
This results both from the more quiescent hydrodynamics in the regions
where the neutrinos were predominantly emitted and from the
multi-angle character of the specific intensity.  Because the local
radiation field is an integral over contributions from many sources at
depth, fluctuations tend to be smoothed out.  This is an important
effect that radial, ``ray-by-ray'' transport methods neglect.

We also extend \cite{OttEtAl08} to investigate the relationship
between variations in the hydrodynamics at depth and observable
variations in the neutrino signal.  Following the calcluations of
\cite{MarekEtAl09} and the detailed post-processing of
\cite{LundEtAl10}, we then estimate the detectability of these rapid
neutrino fluctuations in the water Cherenkov detectors IceCube and
Super-K.  We find that fluctuations characteristic of convective
overturn and shock oscillations could be observed in our nonrotating
model, but are very weak for our rapidly rotating progenitor.
Finally, we calculate the effects of rotation on the anisotropy of the
observed neutrino signal.  We confirm the basic effects of rotation
detailed in \cite{WalderEtAl05}, whereby centrifugal support
suppresses radiation in the equatorial plane.  We also detail the
evolution of the pole-equator anisotropy and calculate the probability
distribution of observable fluxes.

The paper is organized as follows.  In \S\ref{sec:methods}, we
describe our numerical techniques and precollapse models.  In
\S\ref{sec:angvariation}, we present the spatial variation of matter
and radiation fields, highlighting the spatial uniformity of radiation
due to the location of the decoupling region and the multi-angle
character of the specific intensity.  In \S\ref{sec:crosscor}, we
discuss temporal oscillations of matter and radiation, and present the
phasing of dipolar oscillations in these quantities with the shock
position.  Having established a connection between hydrodynamic
oscillations and neutrino fluctuations, we estimate the detectability
of these fluctuations in \S\ref{sec:detect}.  In \S\ref{sec:lumpi}, we
address the impact of rapid rotation on the detectable signal, finding
an orientation-dependent flux asymmetry of up to a factor of $\sim$3
at early times and light curves that strongly depend on both species and
viewing angle.  We conclude in \S\ref{sec:conclusions}.

\section{Methods} \label{sec:methods}

The numerical methods and initial models used in this work are
described in detail in \cite{OttEtAl08} and references therein.  In
this section, we summarize the important points and refer the reader
to these references for a more thorough discussion.

\subsection{Hydrodynamics}

We use the radiation-hydrodynamics code VULCAN-2D, described by
\cite{LivneEtAl04} and \cite{BurrowsEtAl07c}.  The hydrodynamics
module solves the Newtonian Euler equations with artificial viscosity
in two steps: a Lagrangian step followed by a remapping onto an
arbitrary axisymmetric grid.  We use a polar grid with 120 polar
angles and 230 logarithmically spaced radial zones from 20 to 4000 km.
To avoid small cells (and prohibitive time step restrictions), we
transition to a pseudo-Cartesian grid in the inner 20 km.  We
implement Newtonian gravity with a grid-based Poisson solver, and we
use the finite-temperature nuclear equation of state of
\cite{ShenEtAl98a, ShenEtAl98b}.

\subsection{Radiative Transfer}

VULCAN-2D includes two radiative transfer modules: multi-group
flux-limited diffusion (MGFLD) and a discrete ordinates Boltzmann
solver ($S_n$).  Flux-limited diffusion solves the equations of
radiative transfer in the diffusion approximation, using a
flux-limiter to handle the transition to free-streaming.  The $S_n$
solver discretizes the polar and azimuthal angles $\theta$ and $\phi$
of the specific intensity, preserving its multi-angle character.  It
includes emission, absorption, and isotropic scattering with the
transport cross-section $\sigma^s$, related to the total cross-section
$\sigma^s_T$ by $\sigma^s \equiv (1 - \langle\cos\theta\rangle)
\sigma^s_T$, but neglects $O(v/c)$ terms, including Doppler shifts and
neutrino advection.  We use a constant spacing in $\cos\theta$ and a
variable number of azimuthal angles $\phi$ to tile the sphere as
uniformly as possible.  With eight polar angles ($S_8$), we use a
total of 40 $(\theta, \phi)$ pairs, while $S_{12}$ uses 92 and
$S_{16}$ 162 pairs.  These angles are the possible directions of the
specific intensity and, thus, of radiation transport.

For this study we have employed a hybrid of MGFLD and $S_n$, as
discussed in \cite{OttEtAl08}.  We collapse our progenitor models and
follow them after bounce for the ensuing 160 ms using MGFLD.  We then
transition to $S_n$ by freezing the hydrodynamics long enough to allow
the radiation to reach equilibrium with the matter.  We continue to
use MGFLD in the inner 20 km, where the matter is optically thick to
all species and energy groups and the diffusion approximation is
accurate.  We then restart the hydrodynamics and follow the evolution
of the core using this hybrid approach, with MGFLD deep in the core
and $S_n$ at radii greater than 20 km.  Because the computational cost
of $S_n$ scales as $n^2$, we are only able to run the full
evolutionary calculations with eight polar angles.  We have used
$S_{16}$ to compute snapshots of the radiation field at 160 ms after
bounce.

We perform all radiation transport with three neutrino species and
sixteen energy groups.  The first energy group is at 2.5 MeV, and the
rest are logarithmically spaced from 5 to 220 MeV.  While all
neutrinos participate in neutral current reactions, only $\nu_e$ and
$\overline{\nu}_e$ are subject to charged-current interactions.
Because of their similar (neutral-current) cross-sections, we group
$\nu_\mu$, $\overline{\nu}_\mu$, $\nu_\tau$, and $\overline{\nu}_\tau$
into a single ``species,'' which we designate ``$\nu_\mu$.''

\subsection{Progenitor Models}

We begin our calculations with two 20-$M_\odot$ precollapse models of
\cite{WoosleyEtAl02}.  To study the effects of rapid rotation, we
apply an angular velocity profile to one of our models of the form
\begin{equation}
\Omega(R) = \Omega_0 \frac{1}{1 + (R/A)^2},
\end{equation}
where $R$ is the cylindrical radius, $A$ sets the scale of
differential rotation, and $\Omega_0$ is the central angular
frequency.  Because of the computational cost of $S_n$, we are unable
to run a set of models to explore $A$-$\Omega_0$ space.  Instead, we
study the effects of very rapid rotation using $A = 1000$ km and
$\Omega_0 = \pi~\mathrm{rad\,s}^{-1}$, corresponding to an initial
central period of 2 s.  This central period is faster than those
currently favored for evolved massive stars \citep{MaederMeynet00,
HegerEtAl05}.  An important exception is for collapsar models, which
require very fast pre-collapse stellar rotation rates of $\lesssim$5
seconds in the core \citep{Woosley93}.  Without angular momentum loss,
our rotation rate would result in a ``millisecond-period''
protoneutron star, significantly faster than estimated pulsar birth
spin rates \citep{EmmeringChevalier89, Faucher-GiguereKaspi06,
OttEtAl06}.  Our two models therefore bracket the spin range of
plausible progenitors.

\section{Results: Angular Variations in Radiation and Hydrodynamics} \label{sec:angvariation}

There is good evidence that it is easier to explode a supernova by the
delayed neutrino mechanism in two dimensions than in one
\citep{MurphyBurrows08}.  The extra dimension opens a rich array of
instabilities, increasing the residence time of particles in the gain
region where they undergo net neutrino heating
\citep{MurphyBurrows08}.  \cite{NordhausEtAl10} have recently extended
this result to three dimensions, showing that a 3D explosion requires
$\sim$15--25\% less neutrino luminosity than its 2D analogue.  The
central importance of neutrino heating in multiple dimensions
highlights the need to accurately capture the coupling between matter
and radiation in the semi-transparent gain region, over which the
neutrino optical depths range from a few hundredths to a few tenths.

Radiative transfer in the core-collapse supernova context is
fundamentally a seven-dimensional problem, with six dimensions of
phase space and one of time.  In addition, there are six neutrino
species (three particles and their antiparticles).  Because the
neutrinos are not in thermal equilibrium with their surroundings,
simple approximations to full Boltzmann transport may miss important
physical effects.  The most obvious example is the phenomenon of net
neutrino heating, which occurs predominantly because the ambient
radiation field in the region behind the stalled shock is much harder
than it would be in local thermodynamic equilibrium.  In this section,
we present results from our nonrotating model, demonstrating the
striking spatial uniformity of radiation relative to the matter fields
throughout the gain region.  The radiation fields in our rotating
model vary smoothly from pole to equator.  Because the rapid rotation
inhibits convection so strongly, there is little small-scale variation
in either the radiation or the hydrodynamics.  

\subsection{Variation of Neutrino and Matter Fields with Angle} \label{sec:angdep}

The post-shock region of a core-collapse supernova displays a range of
hydrodynamic phenomena, from large-scale shock oscillations and
convective overturn to smaller scale turbulence.  While these
phenomena directly influence the local properties of matter, their
effects on neutrinos are more subtle and depend on the strength of the
coupling between matter and radiation.  In the optically thick core,
the diffusion limit obtains, and radiation and matter are in local
thermodynamic equilibrium.  In the free-streaming limit, the neutrino
properties largely reflect those of matter near the appropriate
neutrinospheres where they are emitted.  We define the
neutrinospheres to be the (energy and species-dependent) radii
$r_\tau$ where the radial optical depths equal $\frac{2}{3}$:
\begin{equation}
\int_{r_\tau}^\infty \langle \kappa (r, \theta) \rho (r, \theta)
\rangle_\Omega \ud r = \frac{2}{3}.
\label{eq:nusphere}
\end{equation}
Because of the rapid transition to free-streaming, the above picture
suggests two relatively distinct regimes.  The angular distribution of
radiation should resemble that of the local matter only where the
material is optically thick.

\begin{figure*}
\includegraphics[height=\figheight]{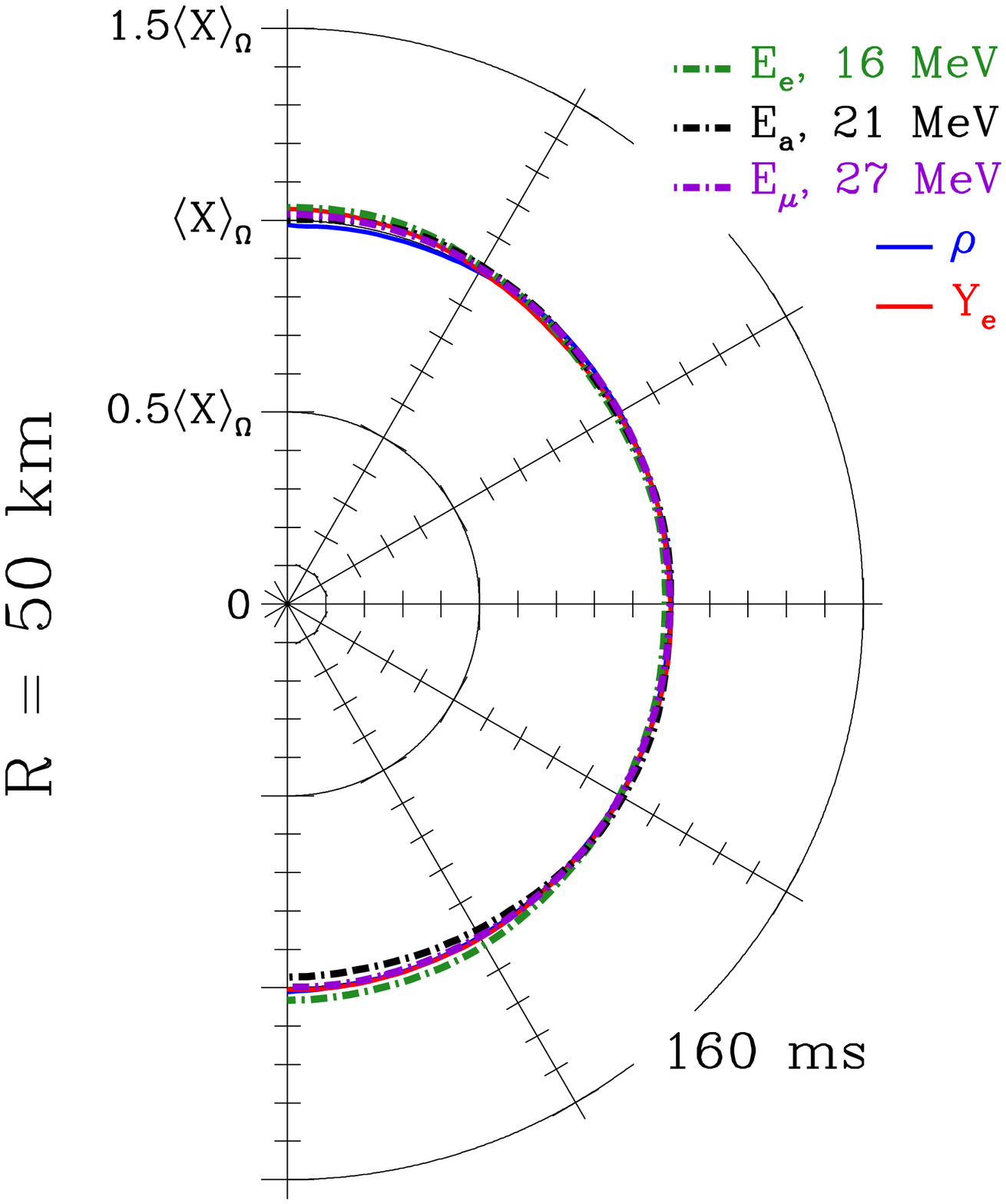}
\includegraphics[height=\figheight]{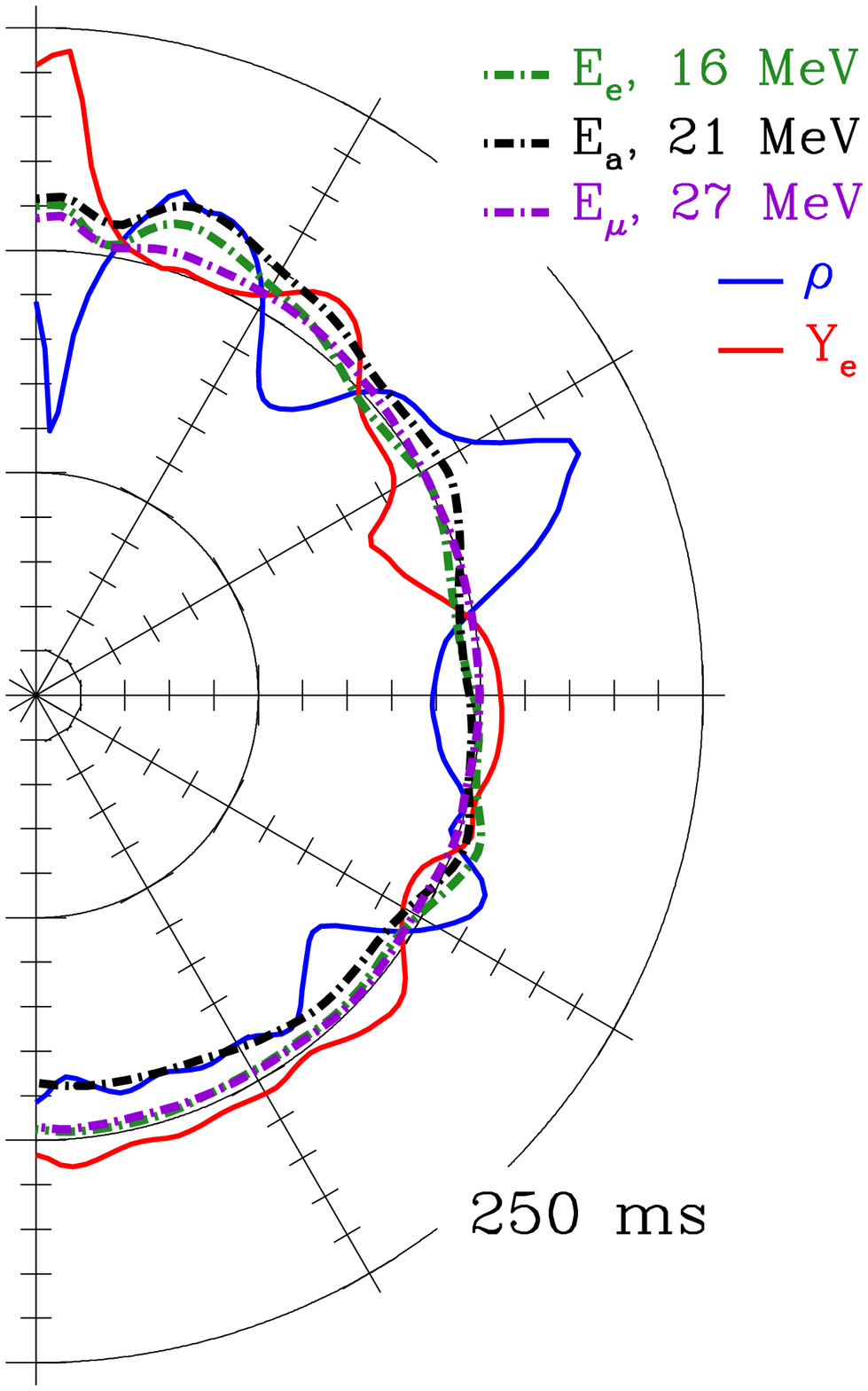}
\includegraphics[height=\figheight]{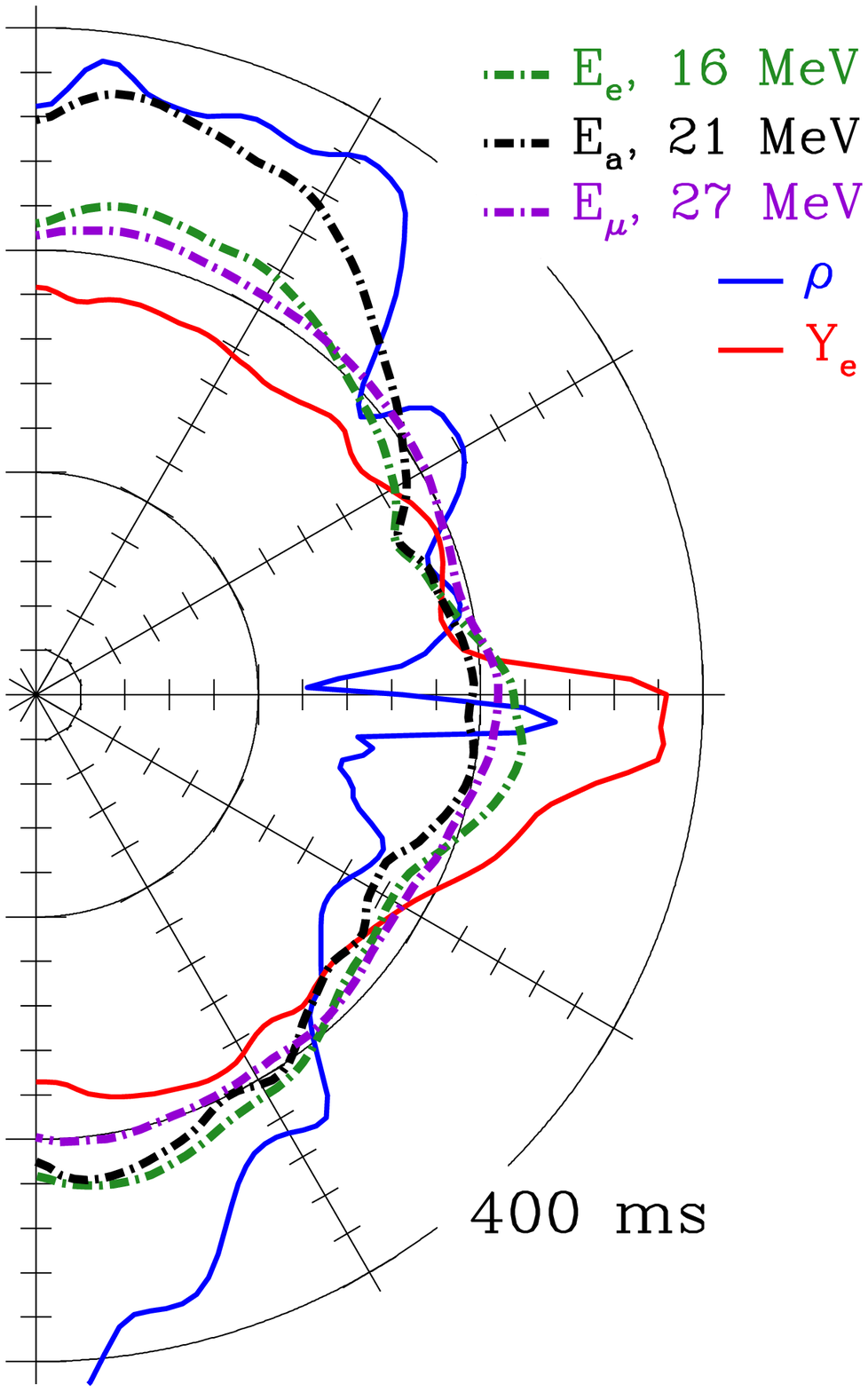}

\includegraphics[height=\figheight]{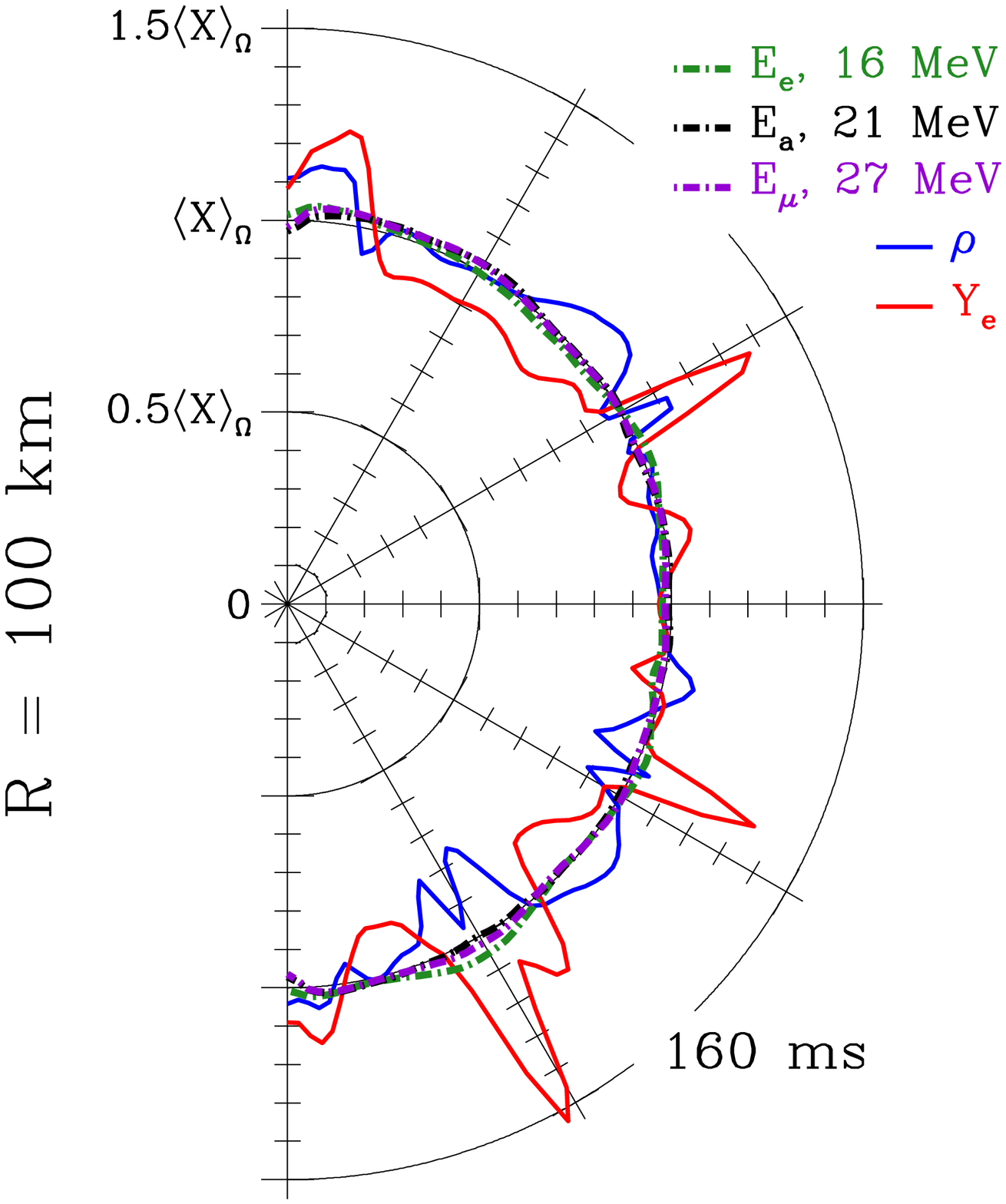}
\includegraphics[height=\figheight]{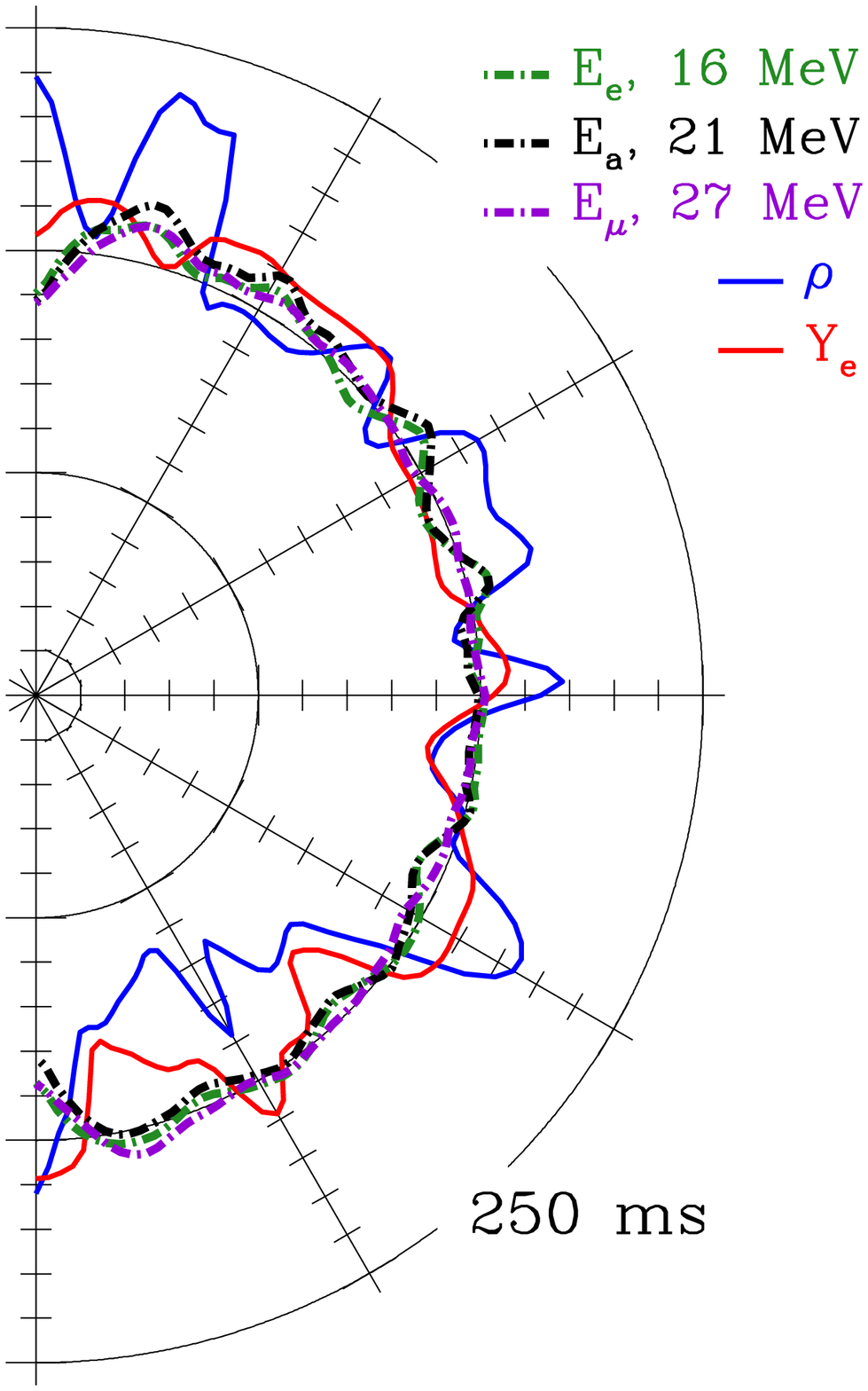}
\includegraphics[height=\figheight]{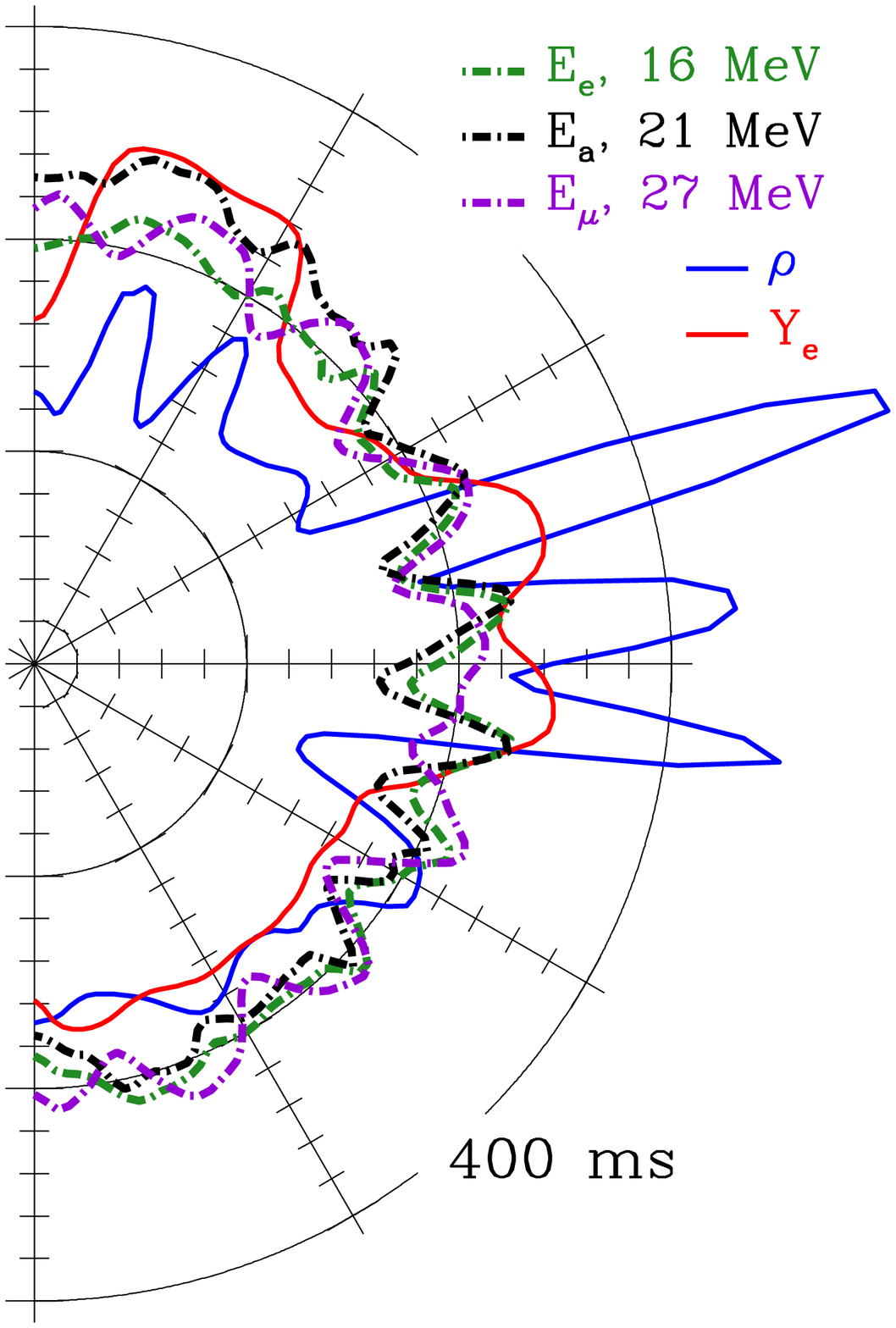}

\caption{Polar plots showing the values of $\rho$, $Y_e$, and the
  spectral energy density of each species near its rms energy
  ($\sqrt{\langle \varepsilon_\nu^2 \rangle}$, Eq.~\ref{eq:nu_rms})
  for the nonrotating model.  $\mathrm{E}_\mathrm{e}$,
  $\mathrm{E}_\mathrm{a}$, and $\mathrm{E}_\mu$ denote electron,
  anti-electron, and mu/tau neutrino types, respectively.  The radial
  coordinate is proportional to each quantity's value at fixed
  physical radius, 50 km in the upper plots and 100 km in the lower
  plots, normalized to its average over $4\pi$ steradians.  The matter
  and radiation fields both show more variation with angle at late
  times, as the shock oscillations grow in amplitude.  However, the
  neutrino energy densities always vary less with angle than the
  thermodynamic quantities $\rho$ and $Y_e$.}
\label{fig:polarplots}
\end{figure*}

We present the angular dependence of hydrodynamic and radiation
quantities in Fig.~\ref{fig:polarplots} at radii of 50 km and 100 km
at three epochs after bounce in our nonrotating model.  These plots
clearly indicate the distinction between the two regimes discussed
above; only the top-left panel shows a region where neutrino spectral
energy densities (indicated at representative energies by dot-dashed
lines) resemble the angular distribution of density $\rho$ and
electron fraction $Y_{\rm e}$.  All quantities are plotted at fixed
radius and normalized to their average values over $4\pi$ steradians.
The neutrino energies used are the root-mean-square values for each
species at large radius,
\begin{equation}
\varepsilon_{\rm rms} =
\sqrt{\langle \varepsilon^2 \rangle} \equiv
\left( \frac{ \int \varepsilon^2 F_\nu (\varepsilon) \ud \varepsilon}
{ \int F_\nu (\varepsilon) \ud \varepsilon} \right)^{1/2},
\label{eq:nu_rms}
\end{equation}
where $F_\nu(\varepsilon)$ is the neutrino energy flux spectrum.  These
rms energies increase with time (see \citealt{OttEtAl08}, Fig.~18).
We use the energy groups closest to the rms values over the first few
hundred milliseconds after bounce: 16, 21, and 27 MeV for $\nu_e$,
$\overline{\nu}_e$, and ``$\nu_\mu$,'' respectively.  Because the
``$\nu_\mu$'' couple to matter more weakly than electron and
anti-electron types, their neutrinospheres lie deeper in the
protoneutron star core.  The ``$\nu_\mu$'' spectra are therefore
harder, reflecting the higher temperatures that prevail where they are
emitted.

As the core evolves during the postbounce phase, variations with angle
increase in all quantities and at all radii.  This is due to two
effects: 1) growth in the vigor of convective overturn and large-scale
shock oscillations, and 2) the contraction of the protoneutron star,
so that a fixed radius moves outward in Lagrangian coordinates.  The
effects of contraction are most clearly seen at a radius of 50 km
between 160 ms (top-left panel of Fig.  \ref{fig:polarplots}) and 250
ms after bounce (top-center panel).  At 160 ms, the base of the
convective zone lies at about 70 km in radius, and the top-left panel
displays the properties of a radiative zone.  By 250 ms after bounce,
the base of the convective zone has sunk to around 45 km.  The region
sampled by the top-center panel is therefore convectively unstable.
As a result, fractional fluctuations in $\rho$ and $Y_e$ increase from
$<$1\% at 50 km and 160 ms after bounce to about 10\% at the same
radius at 250 ms.

While $\rho$ and $Y_e$ vary strongly with angle in the convective
regions, up to a factor of $\sim$2 at late times (see bottom-right
panel of Fig.~\ref{fig:polarplots}), variations in neutrino energy
densities remain smaller throughout the gain region.  Even at 400 ms
after bounce at 100 km (lower-right panel), where the variation in
density reaches a factor of two, fractional variations in neutrino
energy densities remain $\lesssim$20\%.  These neutrinos are
predominately emitted by matter beneath the base of the convective
zone, and retain the relatively smooth angular distribution
characteristic of that region.  As discussed in the next section,
their angular distribution also becomes smoother with increasing
radius due to the integrated contributions from sources at many
angles.  This effect can only be properly captured by full,
multi-angle neutrino transport.  

\subsection{Radial Dependence of Neutrino and Matter Variation} \label{sec:angvar}

Figure \ref{fig:polarplots} indicates a sharp distinction between the
angular dependence of matter in radiative and convective zones.  To
examine the radial dependence of angular variations, we introduce the
fractional angular variance, defined for a quantity X as
\begin{equation}
\sigma^2_\mathrm{X} (r,t) \equiv \frac{ \left<
    \mathrm{X}^2(r,t)\right>_\Omega} {\left< \mathrm{X}(r,t)
    \right>_\Omega^2} - 1,
\label{eq:ang_var}
\end{equation}
where $\langle\mathrm{X} \rangle_\Omega$ denotes the average value of
X over $4\pi$ steradians.  With this definition, $\sigma^2_\mathrm{X}
= 0$ if and only if the quantity X is uniform in angle.  The
fractional angular deviation, $\sigma_\mathrm{X}$, is then
$\sqrt{\sigma^2_\mathrm{X}}$.

\begin{figure*}
\includegraphics[width=0.5\textwidth]{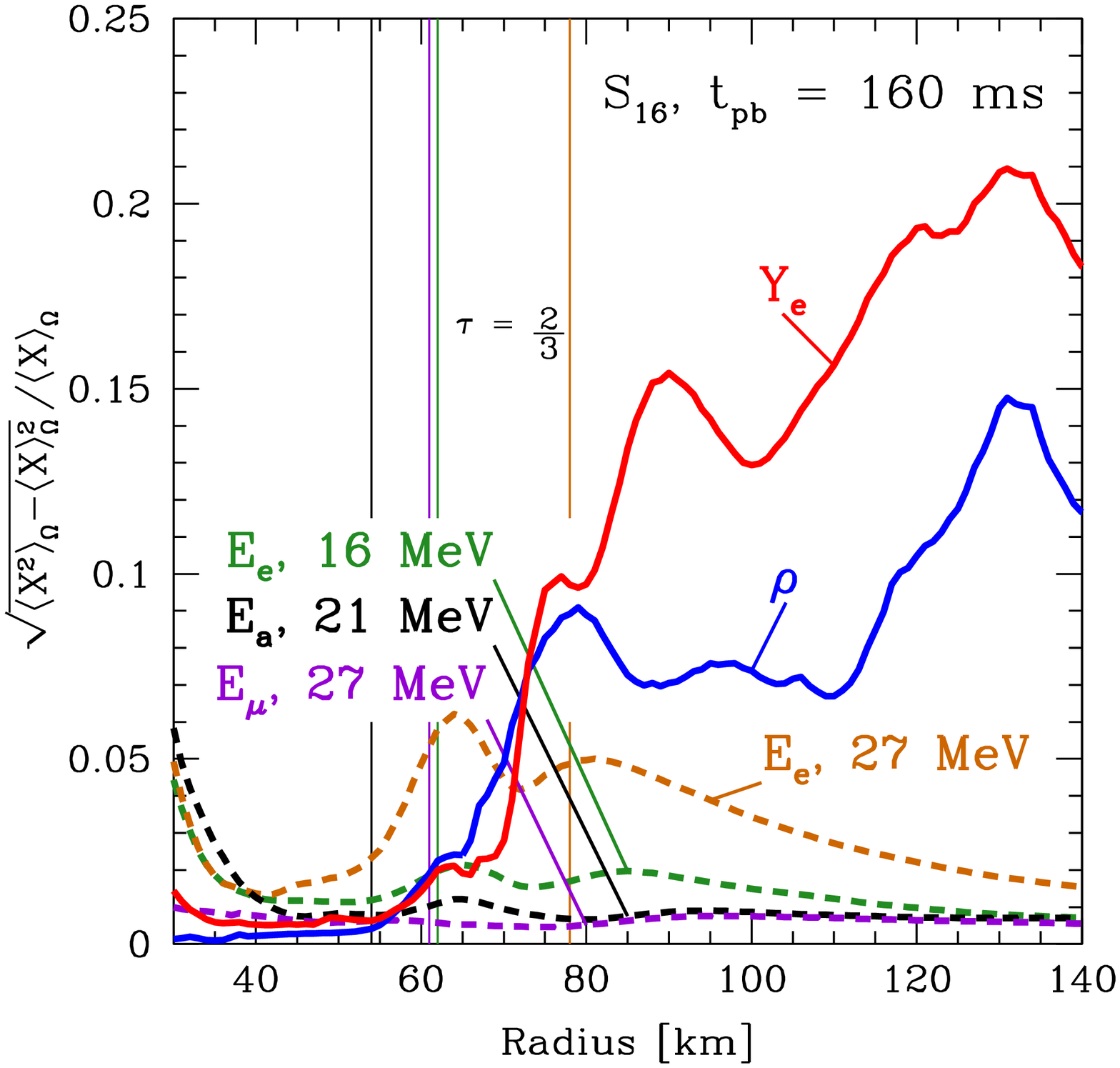}
\includegraphics[width=0.5\textwidth]{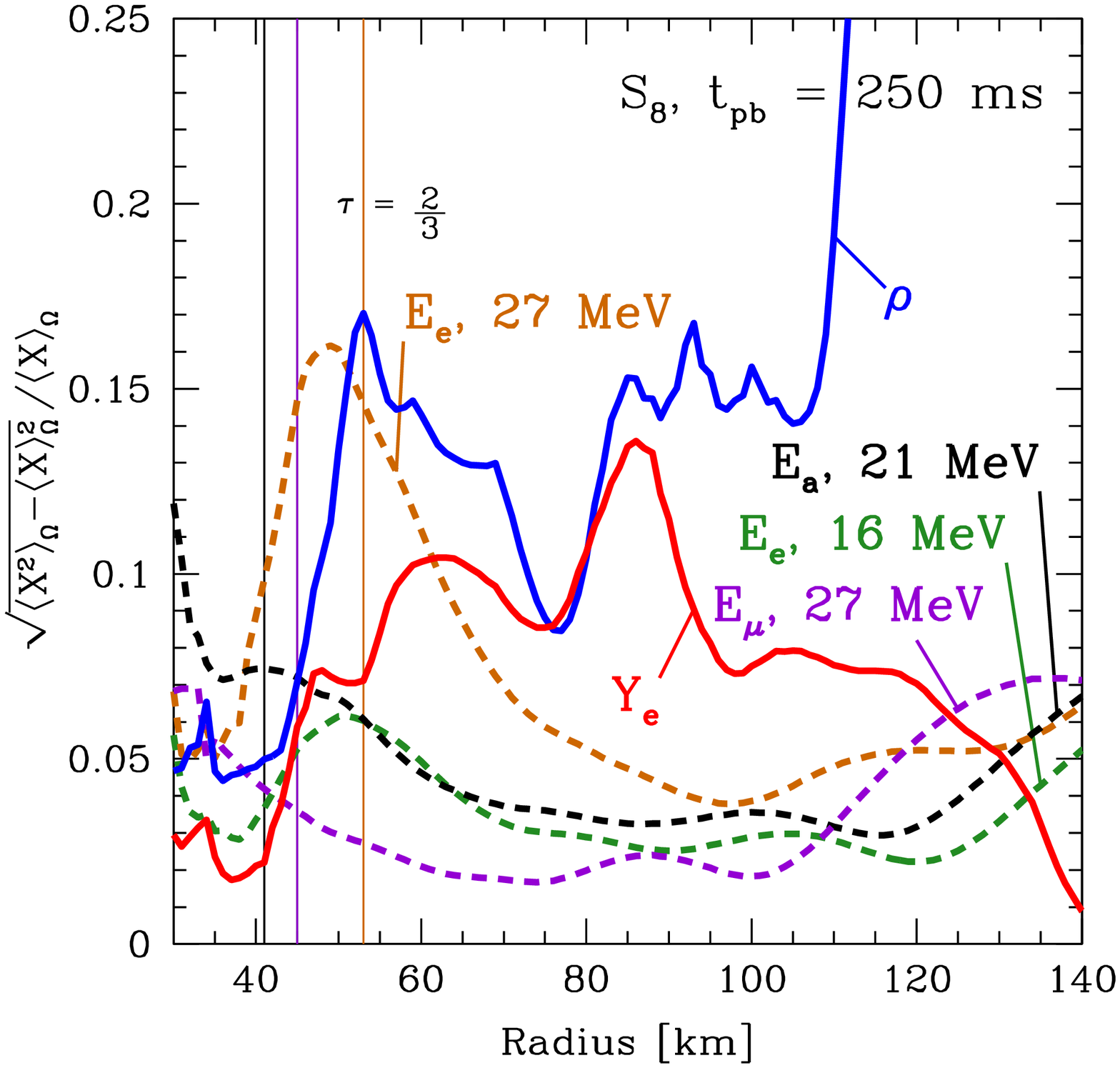}
\caption{Normalized angular deviation ($\sqrt{\langle \mathrm{X}^2
    \rangle_\Omega - \langle \mathrm{X} \rangle_\Omega^2}/\langle
  \mathrm{X} \rangle_\Omega$ in a quantity $\mathrm{X}$,
  Eq.~\eqref{eq:ang_var}, as a function of radius.  Here, $\nu_e$
  represents the spectral energy density (${\rm E}_i$) of $\nu_e$,
  $\mathrm{E}_\mathrm{a}$ of $\overline{\nu}_e$, and $\mathrm{E}_\mu$
  of ``$\nu_\mu$.''  Deviations in $\rho$ and $Y_e$ increase sharply
  at the base of the convective zone and remain high throughout the
  gain region.  Deviations in neutrino energy densities decrease
  beyond the neutrinospheres (denoted by thin vertical lines) due to
  the integrated contribution of sources from many angles at depth.
  Throughout the convective region, angular deviations of energy
  density in all but the highest energy $\nu_e$ are much lower than
  deviations in $\rho$ and $Y_e$.  The neutrino energies shown, except
  27 MeV for $\nu_e$, are each species' approximate rms energy
  ($\sqrt{\langle \varepsilon^2 \rangle}$, Eq.~\eqref{eq:nu_rms}).
  Increases in the angular deviations of $\mathrm{E_i}$ at large
  radius are banding artifacts of the discrete angles used in $S_n$.
  \\}
\label{fig:ang_variance}
\end{figure*}

Figure \ref{fig:ang_variance} shows the fractional angular deviation
of the same quantities plotted in Fig.~\ref{fig:polarplots} at 160 ms
and 250 ms after bounce for our nonrotating model.  The hydrodynamic
variables $\rho$ and $Y_{\rm e}$ clearly show the base of the
convective zone, which sinks from $\sim$70 km at 160 ms to $\sim$45 km
at 250 ms after bounce.  The neutrino spectral energy densities are
relatively uniform throughout the gain region, as suggested by
Fig.~\ref{fig:polarplots}.  This is especially striking at 160 ms
after bounce (left panel), which was computed with 16 polar angles for
neutrino transport.  This snapshot shows increasing angular variation
in the neutrino fields up until roughly the appropriate
neutrinospheres (Eq.~\ref{eq:nusphere}), and then a transition to a
regime where the radiation becomes more spatially uniform with radius.
This is especially apparent in high energy electron neutrinos (like
the 27-MeV energy group shown here).  Because of the $\sim
\varepsilon^2$ dependence of neutrino cross-sections, higher energy
neutrinos have neutrinospheres at larger radii, where the matter begins
to convect.  After the radiation decouples, however, multi-angle
effects smooth its angular distribution.  

Neutrinos in the model at 250 ms after bounce have the same behavior,
first showing more variation with angle and finally becoming smoother
past the neutrinospheres.  However, because only 8 polar angles were
used for neutrino transport, $S_n$ artifacts begin to appear at
$\sim$100--150 km in radius.  In $S_n$, neutrinos can only be
transported along the $n$ discrete polar angles defined in the solver.
In regions of low optical depth, this tends to confine radiation to
radial rays, as discussed in, e.g., \cite{Castor04} and
\cite{LivneEtAl04}.  

The radiation field tends to become more uniform with angle because of
the multi-angle character of the specific intensity; its value at a
point is an integral over contributions from many sources (and
therefore many angles) at depth.  Because an observer at large radius
can effectively ``see'' emission from an entire hemisphere, variations
in the properties of radiation near the neutrinospheres tend to
average out.  As a result, the illumination of matter in the gain
region is more uniform than would be inferred with purely radial
transport.  We quantify this effect in Table \ref{tab:ang_variance},
comparing the fractional angular deviation for neutrino energy
densities at their neutrinospheres to that at 150 km in our
nonrotating $S_{16}$ snapshot.  Table \ref{tab:ang_variance}
demonstrates that the fractional angular deviation is a decreasing
function of radius, as discussed above, but an increasing function of
neutrino energy.  Because higher energy neutrinos decouple at larger
radii, they interact more with convecting matter.  However,
multi-angle effects smooth out much of this variation.  In our
$S_{16}$ snapshot, fractional variations in $\rho$ and $Y_{\rm e}$ are
$\sim$10--20\% throughout the convective zone, while fractional
variations in the neutrino spectral energy densities fall to $\lesssim
2$\% by 150 km.

Many groups currently use ``ray-by-ray'' radial transport\footnote{first 
introduced into supernova theory by \cite{BurrowsEtAl95}} because of
its considerably lower cost \citep[e.g.~][]{BurasEtAl06, MarekJanka09,
  BruennEtAl10}.  The ``ray-by-ray plus'' method \citep{BurasEtAl06}
performs accurate lateral transport in optically thick regions by
keeping terms associated with lateral neutrino advection and pressure
gradients.  However, this technique omits the angular flux terms,
which transport neutrinos relative to the gas, and may therefore
exaggerate the anisotropy of the neutrino distribution at large radius
(c.f.~Section 2.3.2 of \citealt{BurasEtAl06}).  Our results with full
multi-angle transport provide a baseline, which could be used to
calibrate an otherwise ad-hoc coupling of neighboring rays in
semitransparent regions.  Such an approach might provide more accurate
transport at a minimal additional cost over current ``ray-by-ray''
methods.

\begin{deluxetable}{ccccc}
\tablecaption{Angular Deviation in $\nu_e$ Spectral Energy Density}
\tablewidth{0pt}
\tablehead{
\colhead{$\varepsilon_\nu$ [MeV]} & 
\colhead{$r_\tau$ [km]} & 
\colhead{$\sigma (r_\tau)$\tablenotemark{a}} &
\colhead{$\sigma (150~\mathrm{km})$\tablenotemark{b}}
}
\startdata
16 & 62 & 0.020 & 0.0068 \\
21 & 69 & 0.029 & 0.0096 \\
27 & 78 & 0.049 & 0.014 \\
35 & 91 & 0.074 & 0.022 
\enddata 

\tablenotetext{a}{Fractional angular deviation in spectral energy
  density (Eq.~\ref{eq:ang_var}) at the appropriate neutrinosphere
  (Eq.~\ref{eq:nusphere}).}

\tablenotetext{b}{Fractional angular deviation in spectral energy
  density (Eq.~\ref{eq:ang_var}) at 150 km.}

\label{tab:ang_variance}
\end{deluxetable}

\section{The Relationship of Neutrinos to Shock Oscillations} \label{sec:crosscor}

The most visually striking feature of two-dimensional core-collapse
simulations is the large-scale oscillation of matter behind the
stalled shock.  In our nonrotating model, the shock position has a
strong dipolar component that oscillates at $\sim$50 Hz.  In our
rapidly rotating model, the oscillations are much weaker and emerge
several hundred milliseconds later.  The relationship between the
shock position and oscillations in the post-shock material is complex
and has been discussed at length in \cite{FoglizzoTagger00},
\cite{BlondinEtAl03} and \cite{FoglizzoEtAl07}.  Here, we compute the
phasing of oscillations in hydrodynamic and radiation quantities as a
function of radius, tracing the response of matter and radiation
throughout the post-shock region.  We thus establish the temporal
relationship of the observable neutrino emission to the unobservable
hydrodynamics in the core.

We seek to isolate the strongest oscillatory components of the shock
radius and post-shock hydrodynamic and radiation quantities.  Because
these oscillations are dominated by low-$l$ (in particular $l = 1$)
modes in 2D, we use the dipole moments of all quantities.  We define
these to be the $l = 1$ spherical harmonic coefficients in axisymmetry
($m = 0$),
\begin{equation}
a_1({\rm X}(r)) \equiv \frac{\sqrt{3}}{2} \int_0^{2\pi} \ud \phi
\int_0^\pi \mathrm{X}(r, \theta) \cos\theta \sin\theta \ud \theta
\label{eq:dipole}
\end{equation}
in a quantity X.  We also define the normalized cross correlation
function between two quantities X and Y with temporal offset $\tau$ as
\begin{equation}
      C(\mathrm{X}, \mathrm{Y}, \tau) \equiv \frac{\int \ud t
        \left[\mathrm{X}(t)-\left<\mathrm{X} \right> \right] \left[
          \mathrm{Y}(t-\tau) - \left< \mathrm{Y} \right>
          \right]}{\sigma_\mathrm{X} \sigma_\mathrm{Y} \Delta t}.
\label{eq:crosscor}
\end{equation}
Here, $\left< {\rm X} \right>$ denotes the temporal average of X, and
$\sigma_{\rm X}$ its standard deviation.  Thus, $C(\mathrm{X},
\mathrm{X}, 0)$ is the normalized autocorrelation with zero offset and
equals unity, while a negative value of $C$ indicates an
anticorrelation.  We define the delay between X and Y to be the offset
$\tau$ that maximizes their cross-correlation function.  We then
convert this delay into a phase difference using a periodicity of
19.4 ms.  This is the period of the autocorrelation function of the
shock position, $C(R_{\rm sh}, R_{\rm sh}, \tau)$, and represents an
average SASI frequency in our nonrotating model.  In the following
analysis, we use the dipole coefficients $a_1$ of radiation
and hydrodynamic quantities as the inputs to Eq.~\eqref{eq:crosscor}.

\begin{figure}
\includegraphics[width=\linewidth]{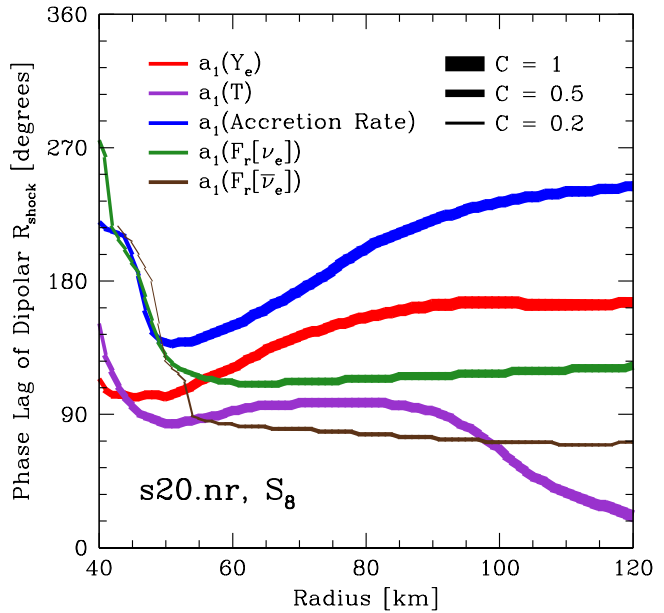}
\caption{Phase (in degrees) by which the dipole component $a_1$ of the
  shock position (Eq.~\ref{eq:dipole}) lags dipolar radiation and
  hydrodynamic quantities.  The line thicknesses are proportional to
  the values of the normalized cross-correlation, with thicker lines
  indicating closer relationships.  The phase lag between the shock
  position and the mass accretion rate and $Y_{\rm e}$ gradually
  decreases, reflecting the fact that the post-shock region is not
  moving as a solid body.  Temperature responds to compression and
  expansion; its maximum asymmetry occurs when the shock position is
  most symmetric (and its velocity is least symmetric).  The phasing
  of neutrino emission is determined near a radius of 50 km, and its
  correlation with the shock position is largest in $\nu_e$.  Electron
  neutrinos have the most emission in convective regions and the
  strongest link to turbulent hydrodynamics. }
\label{fig:crosscor}
\end{figure}

In Fig.~\ref{fig:crosscor}, we show the phase differences between
$a_1(R_{\rm sh})$ and the dipole components of hydrodynamic and
radiation quantities in our nonrotating model.  The vertical axis is
the phase in degrees by which $a_1(R_{\rm sh})$ lags a given quantity,
while the line thickness corresponds to the maximum magnitude of the
cross-correlation function, with thicker lines indicating stronger
correlations.  The hydrodynamic quantities show gradual phase shifts
throughout the post-shock region, while the radiation quantities
display a constant phase past their decoupling radii.  These phase
shifts may be thought of as sonic delays, reflecting the fact that the
post-shock material is not moving as a solid body.

The phase shifts of individual hydrodynamic quantities provide
insights into the dynamics of the post-shock material.  The accretion
rate is closely related to the velocity, and its dipole component is
slightly more than $90^\circ$ out of phase with that of the shock
position.  At the shock radius itself, the phase difference would be
exactly $90^\circ$, as with a simple harmonic oscillator.  The phase
lag varies with depth due to the finite sound speed, reaching an
offset of $180^\circ$ near 70 km.  This delay is also illustrated by
the advected quantity $Y_{\rm e}$.  At large radius, the electron
fraction peaks on the side opposite to that of the shock position due
to the expansion of deleptonized material.  The peak temperature
asymmetry at depth leads the maximum extent of the shock by
$\sim$90$^\circ$.  At large radius, the temperature asymmetry reflects
the shock asymmetry, as the hot post-shock material expands into
unshocked infalling matter.  We have not shown the total heating rate
integrated over the gain region, which is primarily a function of the
volume of the gain region.  Its dipole moment is almost perfectly in
phase with that of the shock radius.

Shock oscillations are hydrodynamic phenomena, and the relationship of
the shock asymmetries to the radiation field is weaker than that to
hydrodynamic quantities.  As shown in Fig.~\ref{fig:crosscor},
however, a clear correlation between the shock radius and neutrino
asymmetries is present, particularly in electron neutrinos.  The
phasing of this relationship is determined near 50 km, and is largely
due to the minority of electron neutrinos emitted from convective
regions.  Anti-electron neutrinos, which decouple almost entirely at
greater depth, display a weaker cross-correlation with dipolar
asymmetries in the shock radius.  The asymmetry in both species'
asymptotic flux is nearly in phase with matter outflows at large
radius, and leads the asymmetry in the shock position by about
$70^\circ$ ($\overline{\nu}_e$) and $120^\circ$ ($\nu_e$).
Though the fractional asymmetries are lower in radiation than in
hydrodynamic quantities, Fig.~\ref{fig:crosscor} demonstrates that
large-scale oscillations are largely responsible for the asymmetries
that we do observe.  A detection of neutrino fluctuations in a real
supernova would provide strong evidence for shock oscillations.

\section{Detectability of Rapid Neutrino Fluctuations} \label{sec:detect}

In the previous section we showed that large-scale oscillations of
matter behind the stalled shock are correlated with oscillations in
neutrino flux, particularly in $\nu_e$'s.  The detectability of these
rapid neutrino fluctuations depends on the fraction of neutrinos
emitted from convective regions and on the vigor of large-scale
overturn and global oscillations.  The fluctuations must be
sufficiently large to be distinguishable from shot noise in a smoothly
declining signal.  Large-scale hydrodynamic oscillations impose a
periodicity on these fluctuations, opening up a variety of
observational tests.  In this section, we examine the prospects for
detecting rapid fluctuations in the emission of a real supernova.  A
robust detection of periodicity in the neutrino signal might confirm
the large-scale oscillations predicted by 2D core-collapse
simulations.

\subsection{Estimating the Signal}

The neutrino signal detected on Earth is the product of the spectral
flux and the detector response function.  For a nearby supernova,
the best counting statistics will be provided by IceCube and Super-K,
both water Cherenkov detectors primarily sensitive to anti-electron
neutrinos.  Each has a response approximately proportional to the
square of the neutrino energy, so that the
\begin{equation}
``\mathrm{signal}" \propto \int_0^\infty f_\varepsilon(\theta, R, t)
  \varepsilon^2 \ud \varepsilon,
\label{eq:signal}
\end{equation}
where the number flux $f$ is a function of viewing angle $\theta$,
supernova distance $R$, and time.  The constant of proportionality
depends on the fiducial volume of the detector, 22.5 kt for Super-K
\citep{IkedaEtAl07} and 940 kt for IceCube \citep{KowarikEtAl09}.
Super-K has almost no background, while IceCube will have an estimated
background rate of approximately $1.34 \times 10^3 ({\rm ms})^{-1}$
\citep{KowarikEtAl09}.  Designed for exceptionally energetic
neutrinos, IceCube would be unable to resolve individual neutrino
energies or trajectories, but would observe a supernova burst as an
increase above its normal background.  Super-K has a negligible
background and would provide approximate neutrino directions and
energies, but with far fewer events.  These two detectors also have
different temporal resolutions.  IceCube bins its data into intervals
of 1.6384 ms \citep{KowarikEtAl09}, while Super-K resolves the
relative arrival time of individual events to microseconds
\citep{IkedaEtAl07}.

\begin{figure}[h]
\includegraphics[width=\linewidth]{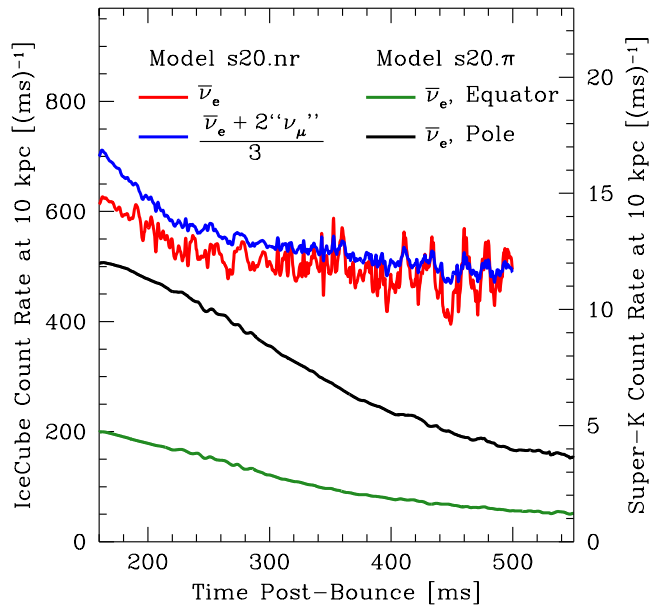}
\caption{Estimated $\overline{\nu}_e$ signals (Eq.~\ref{eq:signal})
  and fully mixed signals $1/3\,\overline{\nu}_e+2/3$``$\nu_\mu$'' of
  our nonrotating model and $\overline{\nu}_e$ ``signals'' of our
  rapidly rotating model.  The fractional fluctuations are as high as
  10\% in the nonrotating model, but $\lesssim$2\% in the rotating
  case.  To the extent that neutrino flavor mixing does occur,
  smoothly declining ``$\nu_\mu$'' will dilute the rapid fluctuations
  shown here in $\overline{\nu}_e$.}
\label{fig:signal}
\end{figure}

In Fig.~\ref{fig:signal}, we show estimated signals computed by
Eq.~\eqref{eq:signal} for both our nonrotating and our rapidly
rotating models.  The fractional fluctuations in electron and
anti-electron estimated signals are $\sim$10\% in the nonrotating
model and are weak functions of viewing angle.  The magnitude of these
fluctuations is much larger than the secular change of the signal over
a SASI period.  In contrast, the estimated signal for our rotating
model is dominated by smoothly declining accretion luminosity and
cooling.  Rapid rotation stabilizes the post-shock region against
convection and suppresses oscillations in the hydrodynamics and
neutrinos.  While these ``signals'' do display some periodic
modulation, the magnitude of rapid fluctuations is comparable to the
secular change over a period.  Orientation effects, which we explore
in detail in \S\ref{sec:lumpi}, dominate in our rapidly rotating
model.

The rapid variations in our neutrino ``signals'' arise from a
combination of vigorous convection, modulated by the SASI, and
neutrino emission from the convecting region.  If the power in shock
oscillations is spread over a wider range of modes (as has been
suggested by recent 3D simulations), these oscillations might leave a
weaker imprint on the neutrinos.  Such a possibility needs to be
addressed quantitatively, ideally with full 3D simulations.
Sophisticated post-processing of 3D hydrodynamic calculations, like
those of \cite{NordhausEtAl10}, may also provide insight.

We also caution that our estimated ``signals'' neglect neutrino flavor
oscillations.  Because they decouple from matter beneath the
convective layer, ``$\nu_\mu$'' show fractional fluctuations of
$\lesssim 1\%$ even in our nonrotating model.  Flavor mixing would
make it more difficult to detect rapid neutrino fluctuations by
diluting the $\overline{\nu}_e$ with these ``$\nu_\mu$.''
\cite{LundEtAl10} found that under an assumption of complete,
energy-independent mixing of all species, fractional fluctuations in
their estimated ``signals'' declined by $\sim$$\frac{2}{3}$.  We also
show detection results for this extreme case in order to bracket the
range of physically plausible neutrino signatures of flux variations,
and find similar results.

\subsection{Detecting the Fluctuations}

The estimated neutrino signals shown in Fig.~\ref{fig:signal} are each
a superposition of a smoothly declining component and a smaller,
rapidly varying component correlated with the hydrodynamics around the
protoneutron star.  Here, we estimate the maximum distance to our
nonrotating supernova model at which these variations would be
detectable by IceCube and Super-K.  \cite{LundEtAl10} estimated the
detectability of rapid fluctuations in their models using
high-frequency Fourier components.  They found that, neglecting flavor
oscillations, IceCube would be able to detect a few high-frequency
components in their model at 10 kpc.  Here, we use the Rayleigh test
\citep{LeahyEtAl83}, a different and simpler method common in radio
and X-ray astronomy, and obtain similar results.

The Rayleigh test is a test for periodicity in a discrete time series.
Given an assumed period, each element of the series is assigned a
phase $\phi$ and thus a unit vector in $r$-$\phi$ space.  The Rayleigh
statistic $\mathcal{R}$ is the normalized magnitude of the vector sum
of these elements,
\begin{equation}
\mathcal{R} = \frac{2}{N}\Bigg[ \bigg( \sum_{i=1}^N \sin \phi_i \bigg)^2 +
  \bigg( \sum_{i=1}^N \cos \phi_i \bigg)^2 \Bigg],
\label{eq:rayleigh}
\end{equation}
over a temporal range consisting of an integer number of periods.  For
a constant signal with only Poisson noise, $\mathcal{R}$ will be be
drawn from a $\chi^2$ distribution with two degrees of freedom
\citep{LeahyEtAl83}.

The Rayleigh test, as described above, takes its null hypothesis to be
a uniform signal.  In the case of supernova neutrino emission, we wish
to extend the null hypothesis to include a smoothly, secularly
changing signal (but still without rapid fluctuations).  We may
accomplish this with a slight modification to the Rayleigh test.  In
addition to the phase angle $\phi$, we define a second phase angle
$\theta \equiv \phi + \pi$.  We take our temporal range for $\theta$
to be an integer number of periods, but with its starting and ending
times offset by one-half period from their values with $\phi$.  In
essence, we use the difference between the first and last half-periods
to calibrate out any secular change.  We then define our modified
Rayleigh statistic $\mathcal{R}'$ to be
\begin{align}
\mathcal{R}' = \frac{4}{N_\phi + N_\theta} \Bigg[ &\frac{1}{4}\bigg(
  \sum_{i=1}^{N_\phi} \sin \phi_i - \sum_{j=1}^{N_\theta} \sin
  \theta_j \bigg)^2 \nonumber \\ &+ \frac{1}{4} \bigg(
  \sum_{i=1}^{N_\phi} \cos \phi_i - \sum_{j=1}^{N_\theta} \cos
  \theta_i \bigg)^2 \Bigg].
\label{eq:rayleigh_mod}
\end{align}
Because of the identities $\sin\theta = -\sin(\theta + \pi)$ and
$\cos\theta = -\cos(\theta + \pi)$, $\mathcal{R}'$ reduces to the
Rayleigh statistic $\mathcal{R}$ except for the different ranges over
which $\theta_i$ and $\phi_i$ are defined.  In the limit of many
detections over many periods, $N_\phi \rightarrow N_\theta$ and
$\mathcal{R}'$ obeys exactly the same statistics as $\mathcal{R}$ but
with a more general null hypothesis.

One may search for periodicity either at a pre-determined frequency or
by sweeping through parameter space.  Here, we perform a coarse sweep,
sampling periods from 5 - 40 ms in intervals of 1 ms.  We choose a
threshold value of 13.16 for $\mathcal{R}'$, which gives a 5\%
probability of a spurious detection.  We then calculate the fraction
of Monte Carlo realizations of signals computed at a given distance
that exceed this threshold in at least one frequency.  These fractions
are our estimated probabilities of detecting high-frequency
periodicity.

\begin{figure}[h]
\includegraphics[width=\linewidth]{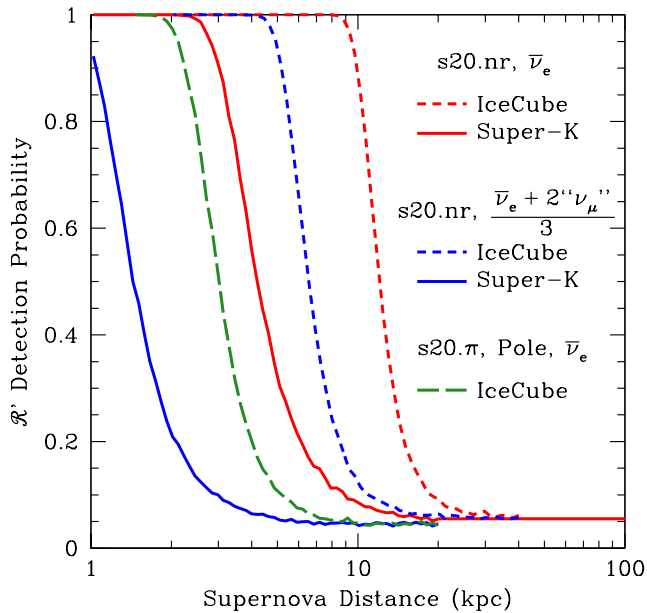}
\caption{The probability of a $2\sigma$ detection of at least one
  rapidly varying component of the neutrino luminosity using the
  modified Rayleigh test (Eq.~\ref{eq:rayleigh_mod}) on Monte Carlo
  realizations of the estimated signals in $\overline{\nu}_e$ and
  $1/3\,\overline{\nu}_e + 2/3$``$\nu_\mu$'' shown in
  Fig.~\ref{fig:signal}.  Depending on the extent of neutrino mixing,
  rapid neutrino fluctuations in our nonrotating model may be
  detectable by IceCube as far away as $\sim$10 kpc.  High frequency
  fluctuations would be detectable by IceCube only within $\sim$3 kpc
  even under the optimistic assumption that no flavor mixing occurs.
  }
\label{fig:pvals}
\end{figure}

Figure \ref{fig:pvals} shows these detection probabilities as a
function of supernova distance for both the IceCube and Super-K
neutrino detectors.  We estimate the detectability under both the
optimistic assumption of no flavor mixing and the pessimistic
assumption that the flavors blend completely.  The true signal will
almost certainly lie somewhere in between, but will require a detailed
calculation with a given mass hierarchy and mixing angle, which is
beyond the scope of this paper.  Even with the pessimistic assumption
of complete flavor interchange, rapid fluctuations in our nonrotating
model would be detectable by IceCube if the supernova were to occur
within $\sim$5 kpc.

\begin{figure}[h]
\includegraphics[width=\linewidth]{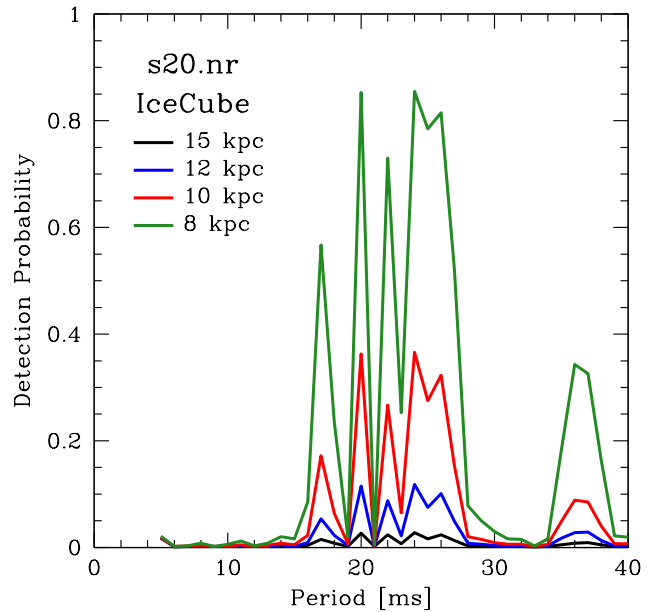}
\caption{The probability of detecting high-frequency variation in the
  anti-electron neutrino luminosity with IceCube and the modified
  Rayleigh test (Eq.~\ref{eq:rayleigh_mod}) at a given periodicity.
  In our nonrotating model, neglecting flavor oscillations, details
  of the SASI should be observable in neutrinos out to $\sim$8 kpc.}
\label{fig:obs_freq}
\end{figure}

Figure \ref{fig:pvals} shows the distance at which {\it any} rapidly
varying component may be detected.  To gain physical insight into the
nature of the shock oscillations, we would need a more detailed power
spectrum of the high-frequency variation.  Figure \ref{fig:obs_freq}
shows the same detection probabilities as plotted for IceCube in
Fig.~\ref{fig:pvals}, but as a function of assumed period at four
distances.  A detailed measurement of the power spectrum of neutrino
fluctuations, with statistically significant detections at several
periods, would be possible for a supernova near the Galactic center,
given the (optimistic) assumption of no flavor mixing.

\section{The Effects of Rotation on Neutrinos} \label{sec:lumpi}

Much of the previous discussion has focused on the consequences of
convection, shock oscillations, and multi-angle neutrino transport.
Our rapidly rotating model is fundamentally different in these
respects.  Convection near the equator is strongly inhibited by
centrifugal support (see the discussion in \citealt{FryerHeger00}).
As shown in Figs. 19 and 21 of \cite{OttEtAl08}, convective overturn
tends to be confined to the polar regions where the centrifugal
support is weaker, and strong shock oscillations do not begin to
develop until about 400 ms after bounce.  The rotating core evolves
relatively smoothly, and is dominated spatially by quadrupole
variations in the matter distribution and neutrino flux.  The
small-scale variations presented in \S\ref{sec:angvariation} and shock
oscillations explored in \S\ref{sec:crosscor} are both strongly
suppressed.

\begin{figure}
\includegraphics[width=\linewidth]{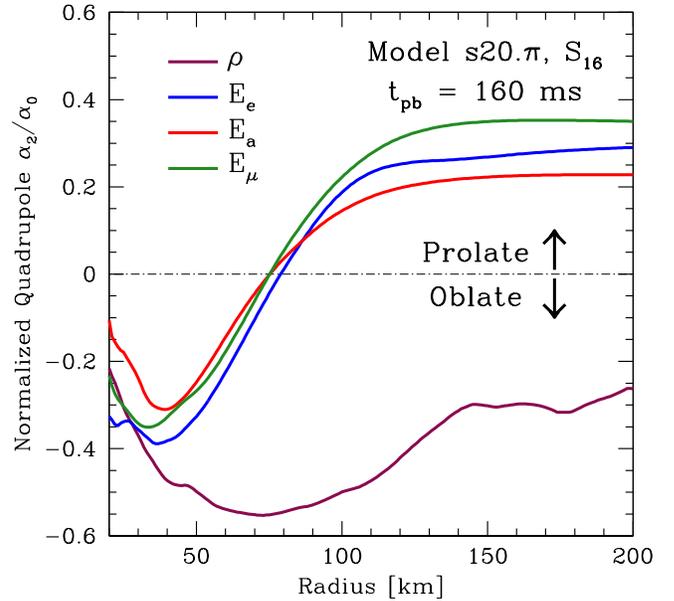}
\caption{Normalized quadrupole moments (Eq.~\ref{eq:quadrupoles}) of
  the neutrino energy densities and matter densities in our rapidly
  rotating model at 160 ms after bounce.  E$_{\rm e}$, E$_{\rm a}$,
  and E$_\mu$ denote the total energy density in $\nu_e$,
  $\overline{\nu}_e$, and ``$\nu_\mu$,'' respectively.  The
  centrifugally-supported matter remains oblate throughout the
  computational domain.  The radiation fields follow the matter in
  optically-thick regions, but become prolate as they decouple, with
  neutrinos escaping more freely near the poles.  }
\label{fig:quadrupoles}
\end{figure}

Rapid rotation creates centrifugal support and an oblate distribution
of matter around the protoneutron star.  At small radii, where the
matter is optically thick, the radiation fields are similarly oblate.
The neutrino distribution transitions to a prolate form beyond the
oblate neutrinospheres \citep{KotakeEtAl03}.  
Figure \ref{fig:quadrupoles} shows this transition.  We measure the
prolateness or oblateness of the density and neutrino energy densities
using the quadrupole coefficient of a spherical harmonic
decomposition.  Normalizing by the average value at (spherical) radius
$r$, we define 
\begin{equation}
\frac{\alpha_2(r)}{\alpha_0(r)} 
\equiv \frac{ \oint d\Omega\, Y_2^0 (\Omega)\,{\rm X}(r)}
{ \oint d\Omega\, Y_0^0 (\Omega)\,{\rm X}(r)}
\label{eq:quadrupoles}
\end{equation}
for a given quantity X.  

The general forms of the matter density and neutrino fluxes, shown in
Fig.~\ref{fig:quadrupoles}, are generic in rotating cores.  The scale
and extent of the matter oblateness are set by the rotation profile.
As discussed in \S\ref{sec:methods} of this paper and in
\cite{OttEtAl08}, rapid rotation in our model produces a relatively
large quadrupole moment in the density and in the asymptotic neutrino
fluxes.  Our model rotates more rapidly than any presented in the
previous studies of \cite{JankaMoenchmeyer89}, \cite{WalderEtAl05},
and \cite{KotakeEtAl03}.  We now examine the effects of such rapid
rotation, from the point of view of an observer meauring the neutrino
flux at a single angle.

Asymmetries in neutrino flux naturally lead a randomly oriented
observer to infer an incorrect luminosity.  We explore and quantify
this effect as a function of neutrino species and of time in our
rapidly rotating model, comparing the inferred luminosity, $4\pi r^2
F_\nu$, to the actual neutrino luminosity,
\begin{equation}
L_\nu = \oint r^2 F_\nu \ud\Omega.
\end{equation}
Here, $r$ represents a radius sufficiently large that all neutrino
species and energies are well into the free-streaming regime.  We use
$r = 150$ km for our calculations in order to satisfy this
requirement, while minimizing the effects of $S_n$ artifacts
(\S\ref{sec:angvar}).

In \S\ref{sec:rotflux1}, we compute the angular distribution of
neutrino flux and the probability distribution of inferred
luminosities in our $S_{16}$ snapshot at 160 ms after bounce.  In
\S\ref{sec:rotflux2}, we show how these distributions evolve in time.
Because of the limited angular resolution of our evolutionary
calculation, $S_n$ artifacts in the asymptotic fluxes become
significant at late times.  We remove this effect at the cost of
angular resolution by integrating over each of the eight bands in
polar angle.

\subsection{Flux Asymmetries: Snapshots} \label{sec:rotflux1}

\begin{figure*}
\includegraphics[width=0.5\linewidth]{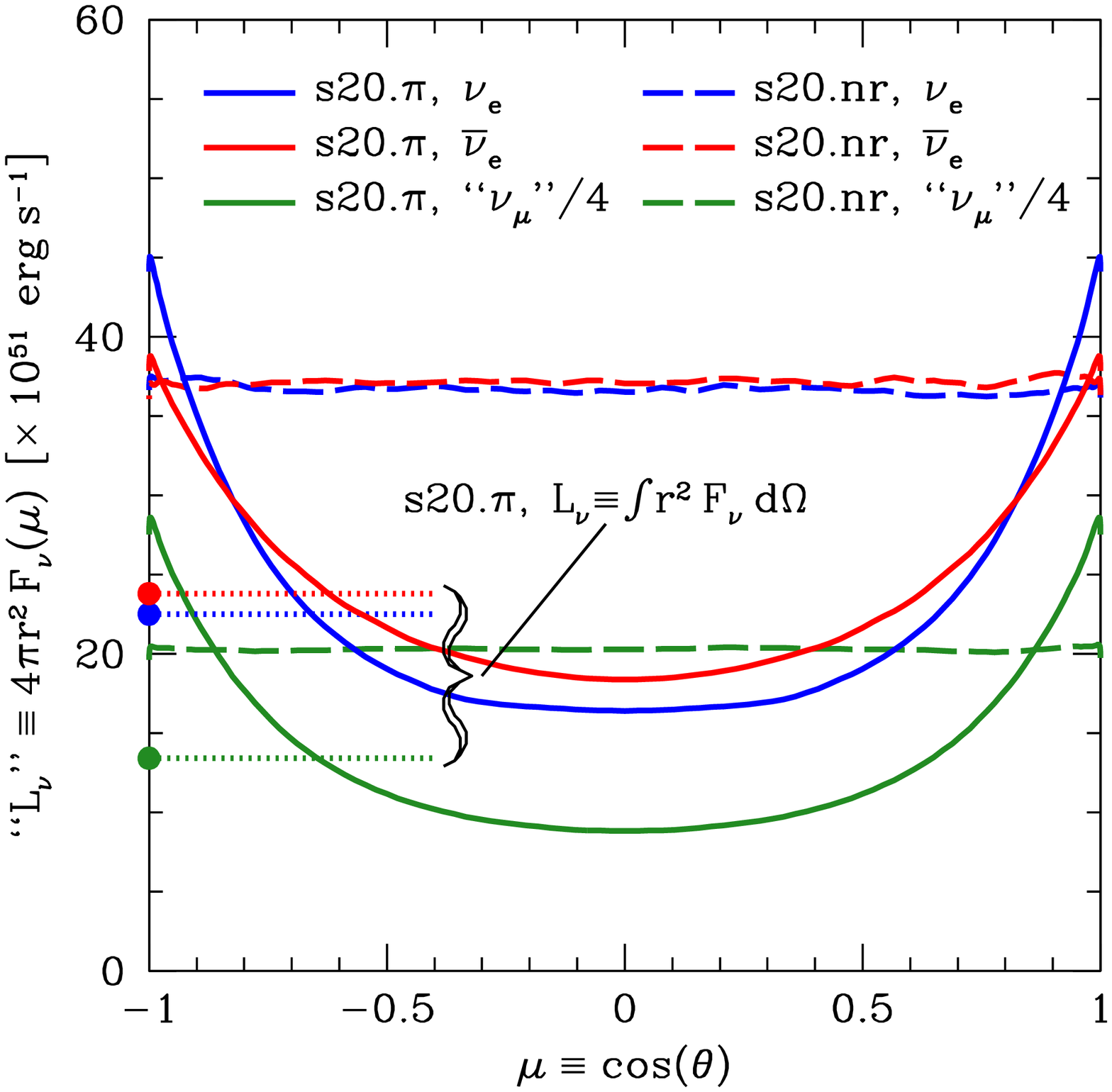}
\includegraphics[width=0.5\linewidth]{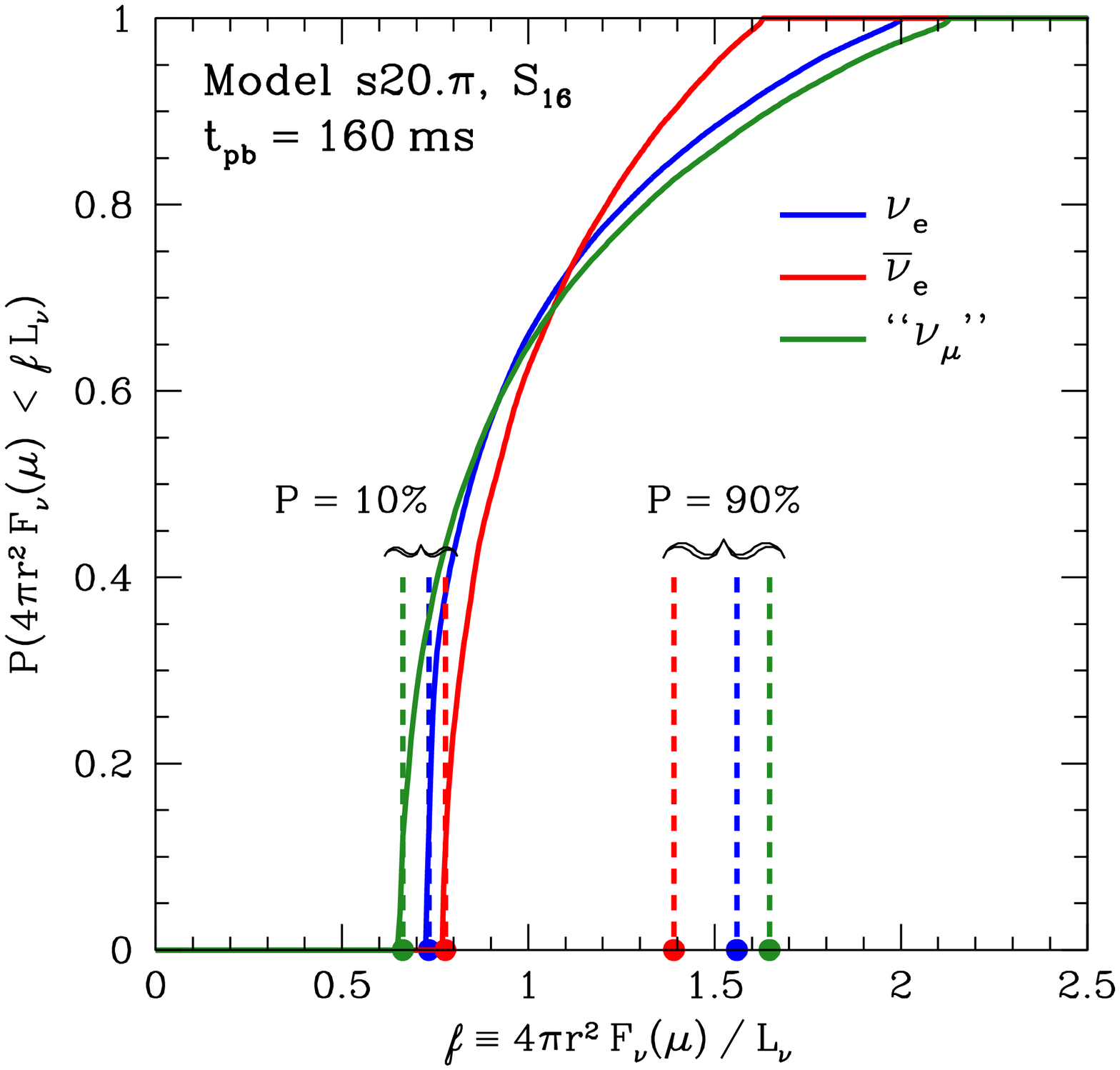}
\caption{{\bf Left panel:} sphericized neutrino luminosities ($4\pi
  r^2 F_\nu$) as a function of viewing angle for the $S_{16}$
  snapshots.  The short dashed lines indicate the true luminosities
  (integrated over $4\pi$ steradians) for the rapidly rotating model.
  Fluxes are calculated at 150 km, after they have nearly reached
  their asymptotic values, but before $S_n$ banding becomes
  noticeable.  {\bf Right panel:} probability of inferring less than a
  given fraction of the true neutrino luminosity in our rapidly
  rotating snapshot.  Even at 160 ms after bounce, there is a 14\%
  chance of overestimating the luminosity in ``$\nu_\mu$'' by at least
  50\% and a 28\% chance of underestimating it by at least 30\%.  This
  is especially important if flavor oscillations mix ``$\nu_\mu$''
  with the more uniform (and more easily detectable) $\nu_e$ and
  $\overline{\nu}_e$.  }
\label{fig:rotation}
\end{figure*}

In Fig.~\ref{fig:rotation}, we present results for our 160 ms $S_{16}$
rotating snapshot.  The left panel shows the total neutrino energy
flux in each species as a function of $\mu \equiv \cos\theta$.  Large
pole-equator flux asymmetries are present in all species, ranging from
a factor of 2.1 in $\overline{\nu}_e$ to a factor of 3.2 for
$``\nu_\mu."$ This corresponds to an overestimate of the true
luminosity by 100\% ($\nu_e$), 60\% ($\overline{\nu}_e$) or 110\%
(``$\nu_\mu$'') at the poles, or an underestimate by 27\% ($\nu_e$),
23\% ($\overline{\nu}_e$), or 34\% (``$\nu_\mu$'') at the equator.
These asymmetries are comparable to or slightly larger than those
reported by \cite{JankaMoenchmeyer89} (who find as much as a factor of
3) and \cite{WalderEtAl05} (who reach a factor of 2.5).  This is
likely due to the extreme rotation of our model, which is higher than
in any of the models presented by \cite{WalderEtAl05}.  In contrast,
pole-equator asymmetries are $\lesssim 4\%$ in the nonrotating model
even though it manifests strong convective plumes and eddies
(cf.~\S\ref{sec:angvariation}).

In the right panel of Fig.~\ref{fig:rotation}, we convert the angular
distribution of neutrino flux into a probability distribution by
assuming the observer to be randomly oriented with respect to the axis
of rotation.  Such an observer would have a 34\% chance of being
within 20\% of the actual luminosity in $\nu_e$, a 54\% chance in
$\overline{\nu}_e$, and a 28\% chance in ``$\nu_\mu$.''  These
distributions have significant tails, particularly for ``$\nu_\mu$''s
for which there is a 14\% chance of observing at least 1.5 times the
mean flux and a 2.3\% chance of observing at least twice the mean.

All of these calculations neglect neutrino flavor oscillations.
Should the species mix, anti-electron neutrinos detectable in IceCube
and Super-K could blend with mu and tau antineutrinos, increasing the
observable asymmetry in $\overline{\nu}_e$.  Flavor oscillations would
thus increase the uncertainty in inferring the true neutrino
luminosity.

\subsection{Light Curve Asymmetries} \label{sec:rotflux2}

A core-collapse supernova's neutrino light curve is a probe of the
physical processes deep within the collapsed core.  There is a burst
of $\nu_e$ emission as the shock wave reaches the electron
neutrinospheres \citep{BurrowsMazurek83}.  The early postbounce phase is powered by accretion
onto the core, while much of the total energy emerges during the tens
of seconds duration of the protoneutron star phase
\citep{BurrowsLattimer86}.  A rapidly rotating supernova
will have a light curve strongly dependent on viewing angle,
particularly during the accretion-powered phase a few hundred
milliseconds after bounce.  In this section, we compute the light curve
shape as a function of viewing angle in both our nonrotating and our
rapidly rotating models.

Unfortunately, our evolutionary calculations lack the polar angle
resolution to accurately calculate the angular distribution of
neutrino fluxes at large radius.  As the neutrinospheres sink and
optical depths decrease at fixed radius, $S_n$ banding becomes
significant at smaller radii.  We remove these artifacts at the cost
of angular resolution, by calculating average fluxes over angular
wedges of $22.5^\circ$ (an $S_n$ band, or one-eighth of a hemisphere).
We may still measure the evolution of the pole-equator asymmetry by
choosing one band near the pole and a second near the equator.
Because the flux is nearly symmetric with respect to the equator
(cf.~Fig.~18 of \cite{OttEtAl08}), we only show results for polar
viewing angles in the range $0^\circ < \theta < 90^\circ$.  We
normalize all observed fluxes to their values at 160 ms after bounce
to show evolutionary differences.  These are arguably more significant
than the total inferred power, as they are insensitive to
uncertainties in distance.

\begin{figure*}
\includegraphics[width=0.5\linewidth]{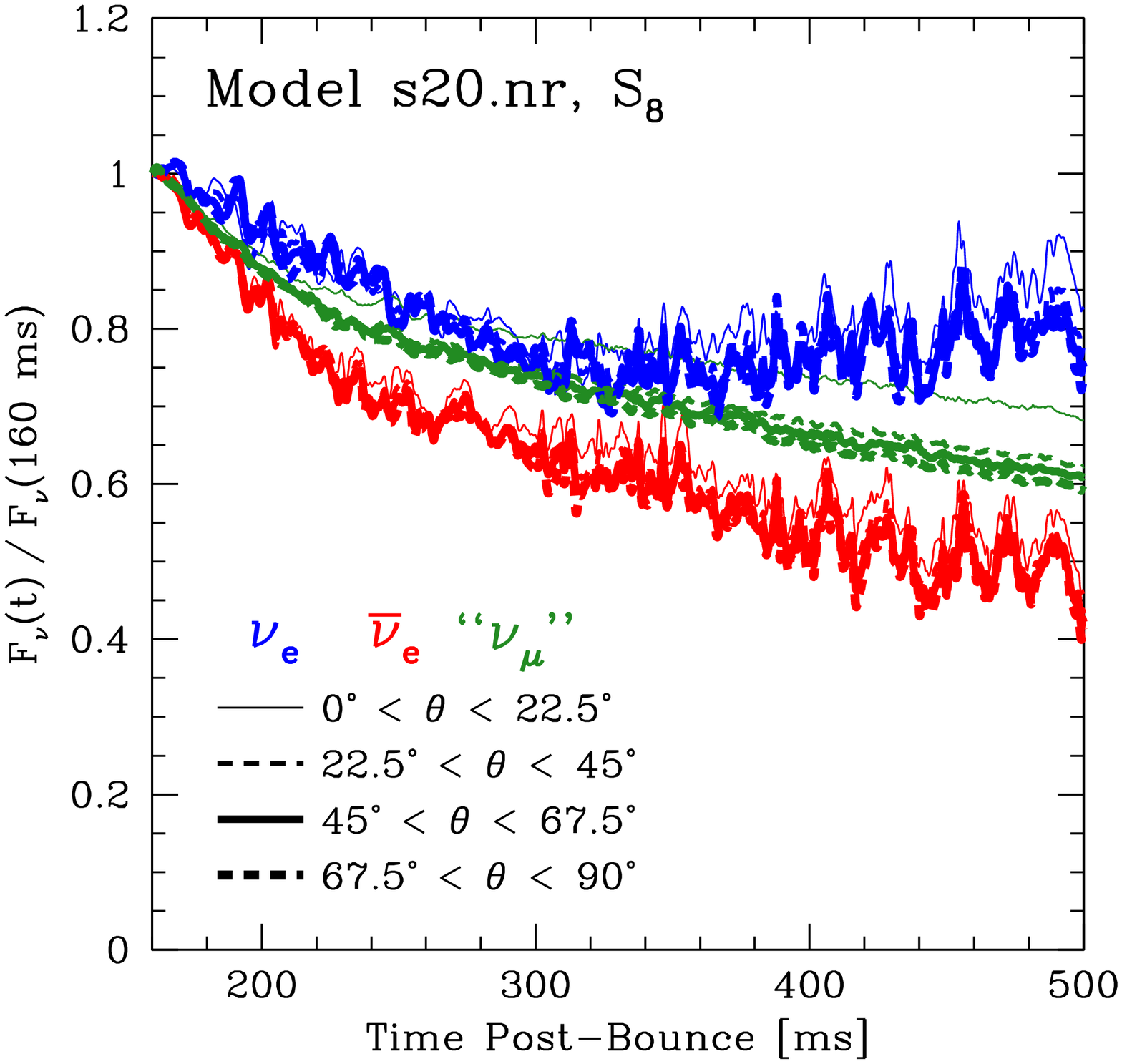}
\includegraphics[width=0.5\linewidth]{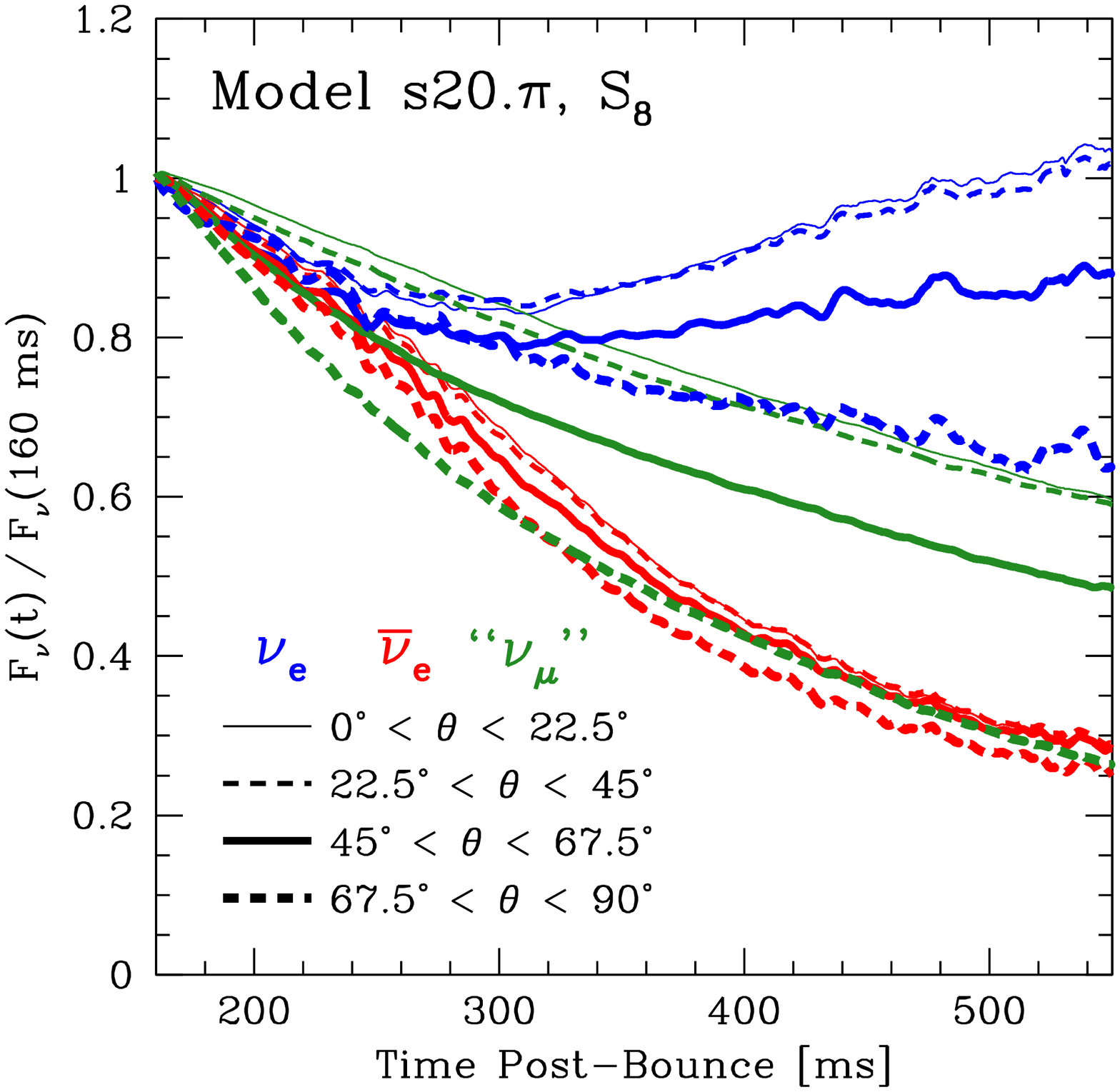}
\caption{Light curves of the nonrotating ({\bf left panel}) and
  rapidly rotating ({\bf right panel}) models over four ranges of
  polar viewing angles from 0 to 90$^\circ$, normalized to 160 ms
  after bounce.  The angular ranges integrate over $S_n$ artifacts,
  and we use line thicknesses proportional to the solid angle
  subtended.  Because the flux is nearly symmetric about the equator,
  we only show viewing angles for the northern hemisphere.  The
  nonrotating model has orientation effects of $\lesssim 10\%$,
  comparable in magnitude to temporal fluctuations.  The rapidly
  rotating model has a nearly constant asymmetry in $\overline{\nu}_e$
  flux, but its ``$\nu_\mu$'' light curve declines twice as fast near
  the equator as it does near the north pole.  The $\nu_e$ flux
  decreases by $\sim$35\% near the equator, while actually {\it
  increasing} near a pole.  Because of the differences between light
  curves of the various species, flavor oscillations could have a
  large effect on the shape of observed light curves.  }
\label{fig:lumevol}
\end{figure*}

We show our results in Fig.~\ref{fig:lumevol}.  The nonrotating
model (left panel) shows little variation in its light curve with
angle.  In this case, variations with angle are $\lesssim 10$\%,
comparable to or lower in magnitude than rapid temporal fluctuations.
However, orientation effects dominate the light curves of the rapidly
rotating model (right panel).  The shape of the light curve shows
little variation with angle in $\overline{\nu}_e$, but declines twice
as much in thermal ``$\nu_\mu$''s near the equator as near a pole.  The
flux in $\nu_e$, while declining by $\sim$35\% near the equator,
actually {\it increases} for an observer near a pole.  Because water
Cherenkov detectors like IceCube and Super-K are primarily sensitive
to $\overline{\nu}_e$, the degree to which the less uniform
``$\nu_\mu$'' mix with $\overline{\nu}_e$ will determine not only the
observed luminosity, but also the shape of the observed neutrino light
curve.  

\section{Conclusions} \label{sec:conclusions}

We have presented new results from the only 2D core-collapse calculations
with full multi-angle neutrino transport.  We have computed the
spatial distributions of matter and radiation fields, finding that the
radiation fields are more uniform than matter fields throughout the
crucial gain region.  This is both a result of the location of the
decoupling region (primarily beneath the base of the convective
region) and due to the multi-angle character of the specific
intensity, so that the radiation at a point is an integral over many
sources at depth.  This latter effect can only be properly captured by
full multi-angle neutrino transport.  Our calculations therefore
provide an important calibration test of methods, like the
``ray-by-ray'' approach, that do not naturally handle lateral
transport.

We have measured the relationship of the radiation fields to the
large-scale, low-order mode shock oscillations, the most striking
feature of 2D core-collapse simulations.  This provides an important
way to connect observable fluctuations in the neutrino luminosity to
the otherwise unobservable hydrodynamics around the collapsed core.
We have also calculated the phasing of the dipole moments of
hydrodynamic quantities with the shock radius, demonstrating both the
temporal structure of the oscillations and their variation with depth.

Following earlier work by \cite{MarekEtAl09} and \cite{LundEtAl10}, we
have estimated the detectability of temporal fluctuations in the
neutrino luminosity by the current water Cherenkov detectors IceCube
and Super-K.  We use a novel variant of the Rayleigh test for
periodicity.  With our modification, the test takes its null
hypothesis to be any smoothly varying signal, not simply a constant
signal.  This modified Rayleigh test applies to a wider range of
problems in which a periodic, rapidly varying component is
superimposed on {\it any} secularly changing signal.  We find that,
neglecting neutrino flavor oscillations, we expect to measure a
reasonably detailed power spectrum with IceCube for our nonrotating
model within $\sim$8 kpc.  Under these same (likely optimistic)
assumptions, Super-K would be able to detect rapid fluctuations out to
a distance of only $\sim$3 kpc.  Given the pessimistic assumption that
all neutrino species mix uniformly, rapid fluctuations in our
nonrotating model would be detectable in IceCube out to $\sim$5 kpc,
but a detailed power spectrum would probably lie beyond the
capabilities of current detectors.  The true signal will almost
certainly lie somewhere in between these extremes of no and total
mixing.  This raises interesting prospects for the next Galactic
supernova, if the low-mode shock oscillations in our nonrotating model
are comparable in magnitude to those found in Nature.  Additional and
complementary information could be gained by the observation of
gravitational waves that carry with them the imprint of
multi-dimensional dynamics in core and postshock region \citep{Ott09,
MurphyEtAl09, MarekEtAl09, YakuninEtAl10}.

We also analyzed the angular distribution of radiation in our rapidly
rotating model and its effects on the luminosities and light curves
measured by randomly oriented observers.  We find pole-equator
asymmetries at least as large as \cite{JankaMoenchmeyer89}, and larger
than any of the models studied by \cite{WalderEtAl05} (though ours
also rotates somewhat faster).  The asymmetries are strongly dependent
on neutrino species, with ``$\nu_\mu$'' showing the greatest asymmetry
(a factor of $\sim$3 at 160 ms after bounce) and $\overline{\nu}_e$ the
least (a factor of $\sim$2).  The light curves of our rapidly rotating
model are also strong functions of species and of angle, and again,
the ``$\nu_\mu$'' show the greatest asymmetry and $\overline{\nu}_e$ the
least.  By 550 ms after bounce, the pole-equator asymmetry remains a
factor of $\sim$2 in $\overline{\nu}_e$, but has grown to $\sim$6 in
``$\nu_\mu$.''  The asymmetry in any observed signal thus depends
strongly on the degree of neutrino flavor mixing.  Because of the
strong dependence of the light curves on angle, a randomly oriented
observer could face considerable uncertainty in the timescale, not
simply the magnitude, of the neutrino emission.

By treating the neutrino-matter coupling accurately in the crucial
gain region, our calculations have enabled a detailed study of the
spatial and temporal character of the radiation fields.  These will
enable the calibration of less accurate (and much less expensive)
radiative transfer techniques.  They also demonstrate the strikingly
uniform character of radiation in the gain region.  Our estimates of
the detectable features of our models also suggest that the next
Galactic supernova may offer an observational test of the dipolar
shock oscillations common to 2D core-collapse simulations, and
highlight the uncertainty associated with neutrino measurements of a
rapidly rotating core.

\begin{acknowledgments}
The authors acknowledge fruitful ongoing and past collaborations with,
conversations with, or input from Jason Nordhaus, Emmanouela Rantsiou,
and Evan O'Connor. This material is based upon work supported under a
National Science Foundation Graduate Research Fellowship to TDB.
A.B. is supported by the Scientific Discovery through Advanced
Computing (SciDAC) program of the DOE, under grant number
DE-FG02-08ER41544, the NSF under the subaward no. ND201387 to the
Joint Institute for Nuclear Astrophysics (JINA, NSF PHY-0822648), and
the NSF PetaApps program, under award OCI-0905046 via a subaward
no. 44592 from Louisiana State University to Princeton University.
C.D.O. is partially supported by the NSF under grant numbers
AST-0855535 and OCI-0905046.  Computational resources were provided by
the TIGRESS high performance computer center at Princeton University,
which is jointly supported by the Princeton Institute for
Computational Science and Engineering (PICSciE) and the Princeton
University Office of Information Technology.  Other computational
resources used include the NSF TeraGrid under award TG-PHY100033, the
Caltech NSF MRI-R$^2$ cluster Zwicky (PHY-1057238), and the Louisiana
Optical Network Infrastructure compute clusters under award
loni\_numrel05.
\end{acknowledgments}

\bibliographystyle{apj_eprint}
\bibliography{Sn}

\begin{thebibliography}{}

\bibitem[\protect\citeauthoryear{{Bethe}}{{Bethe}}{1990}]{Bethe90}
{Bethe}, H.~A. 1990, Reviews of Modern Physics, 62, 801

\bibitem[\protect\citeauthoryear{{Bethe} \& {Wilson}}{{Bethe} \&
  {Wilson}}{1985}]{BetheWilson85}
{Bethe}, H.~A.,  \& {Wilson}, J.~R. 1985, \apj, 295, 14

\bibitem[\protect\citeauthoryear{{Blondin}, {Mezzacappa}, \&
  {DeMarino}}{{Blondin} et~al.}{2003}]{BlondinEtAl03}
{Blondin}, J.~M., {Mezzacappa}, A.,  \& {DeMarino}, C. 2003, \apj, 584, 971

\bibitem[\protect\citeauthoryear{{Bruenn} et~al.}{{Bruenn}
  et~al.}{2010}]{BruennEtAl10}
{Bruenn}, S.~W., {Mezzacappa}, A., {Hix}, W.~R., {Blondin}, J.~M.,
  {Marronetti}, P., {Messer}, O.~E.~B., {Dirk}, C.~J.,  \& {Yoshida}, S. 2010,
  ArXiv e-prints, 1002.4914

\bibitem[\protect\citeauthoryear{{Buras} et~al.}{{Buras}
  et~al.}{2006}]{BurasEtAl06}
{Buras}, R., {Rampp}, M., {Janka}, H.,  \& {Kifonidis}, K. 2006, \aap, 447,
  1049

\bibitem[\protect\citeauthoryear{{Burrows}, {Dessart}, \& {Livne}}{{Burrows}
  et~al.}{2007}]{BurrowsEtAl07a}
{Burrows}, A., {Dessart}, L.,  \& {Livne}, E. 2007, in American Institute of
  Physics Conference Series, Vol. 937, Supernova 1987A: 20 Years After:
  Supernovae and Gamma-Ray Bursters, ed. {S.~Immler, K.~Weiler, \& R.~McCray},
  370

\bibitem[\protect\citeauthoryear{{Burrows}, {Hayes}, \& {Fryxell}}{{Burrows}
  et~al.}{1995}]{BurrowsEtAl95}
{Burrows}, A., {Hayes}, J.,  \& {Fryxell}, B.~A. 1995, \apj, 450, 830

\bibitem[\protect\citeauthoryear{{Burrows} \& {Lattimer}}{{Burrows} \&
  {Lattimer}}{1986}]{BurrowsLattimer86}
{Burrows}, A.,  \& {Lattimer}, J.~M. 1986, \apj, 307, 178

\bibitem[\protect\citeauthoryear{{Burrows} et~al.}{{Burrows}
  et~al.}{2007}]{BurrowsEtAl07c}
{Burrows}, A., {Livne}, E., {Dessart}, L., {Ott}, C.~D.,  \& {Murphy}, J. 2007,
  \apj, 655, 416

\bibitem[\protect\citeauthoryear{{Burrows} \& {Mazurek}}{{Burrows} \&
  {Mazurek}}{1983}]{BurrowsMazurek83}
{Burrows}, A.,  \& {Mazurek}, T.~L. 1983, \nat, 301, 315

\bibitem[\protect\citeauthoryear{{Castor}}{{Castor}}{2004}]{Castor04}
{Castor}, J.~I. 2004, {Radiation Hydrodynamics}

\bibitem[\protect\citeauthoryear{{Colgate} \& {White}}{{Colgate} \&
  {White}}{1966}]{ColgateWhite66}
{Colgate}, S.~A.,  \& {White}, R.~H. 1966, \apj, 143, 626

\bibitem[\protect\citeauthoryear{{Demorest} et~al.}{{Demorest}
  et~al.}{2010}]{DemorestEtAl10}
{Demorest}, P.~B., {Pennucci}, T., {Ransom}, S.~M., {Roberts}, M.~S.~E.,  \&
  {Hessels}, J.~W.~T. 2010, \nat, 467, 1081

\bibitem[\protect\citeauthoryear{{Emmering} \& {Chevalier}}{{Emmering} \&
  {Chevalier}}{1989}]{EmmeringChevalier89}
{Emmering}, R.~T.,  \& {Chevalier}, R.~A. 1989, \apj, 345, 931

\bibitem[\protect\citeauthoryear{{Faucher-Gigu{\`e}re} \&
  {Kaspi}}{{Faucher-Gigu{\`e}re} \& {Kaspi}}{2006}]{Faucher-GiguereKaspi06}
{Faucher-Gigu{\`e}re}, C.,  \& {Kaspi}, V.~M. 2006, \apj, 643, 332

\bibitem[\protect\citeauthoryear{{Fern{\'a}ndez}}{{Fern{\'a}ndez}}{2010}]{Fern%
andez10}
{Fern{\'a}ndez}, R. 2010, ArXiv e-prints, 1003.1730

\bibitem[\protect\citeauthoryear{{Foglizzo} et~al.}{{Foglizzo}
  et~al.}{2007}]{FoglizzoEtAl07}
{Foglizzo}, T., {Galletti}, P., {Scheck}, L.,  \& {Janka}, H. 2007, \apj, 654,
  1006

\bibitem[\protect\citeauthoryear{{Foglizzo} \& {Tagger}}{{Foglizzo} \&
  {Tagger}}{2000}]{FoglizzoTagger00}
{Foglizzo}, T.,  \& {Tagger}, M. 2000, \aap, 363, 174

\bibitem[\protect\citeauthoryear{{Fryer} \& {Heger}}{{Fryer} \&
  {Heger}}{2000}]{FryerHeger00}
{Fryer}, C.~L.,  \& {Heger}, A. 2000, \apj, 541, 1033

\bibitem[\protect\citeauthoryear{{Fryer} \& {Warren}}{{Fryer} \&
  {Warren}}{2002}]{FryerWarren02}
{Fryer}, C.~L.,  \& {Warren}, M.~S. 2002, \apjl, 574, L65

\bibitem[\protect\citeauthoryear{{Fryer} \& {Warren}}{{Fryer} \&
  {Warren}}{2004}]{FryerWarren04}
{Fryer}, C.~L.,  \& {Warren}, M.~S. 2004, \apj, 601, 391

\bibitem[\protect\citeauthoryear{{Fryer} \& {Young}}{{Fryer} \&
  {Young}}{2007}]{FryerYoung07}
{Fryer}, C.~L.,  \& {Young}, P.~A. 2007, \apj, 659, 1438

\bibitem[\protect\citeauthoryear{{Heger}, {Woosley}, \& {Spruit}}{{Heger}
  et~al.}{2005}]{HegerEtAl05}
{Heger}, A., {Woosley}, S.~E.,  \& {Spruit}, H.~C. 2005, \apj, 626, 350

\bibitem[\protect\citeauthoryear{{Herant} et~al.}{{Herant}
  et~al.}{1994}]{HerantEtAl94}
{Herant}, M., {Benz}, W., {Hix}, W.~R., {Fryer}, C.~L.,  \& {Colgate}, S.~A.
  1994, \apj, 435, 339

\bibitem[\protect\citeauthoryear{{Ikeda} et~al.}{{Ikeda}
  et~al.}{2007}]{IkedaEtAl07}
{Ikeda}, M., et~al. 2007, \apj, 669, 519

\bibitem[\protect\citeauthoryear{{Iwakami} et~al.}{{Iwakami}
  et~al.}{2008}]{IwakamiEtAl08}
{Iwakami}, W., {Kotake}, K., {Ohnishi}, N., {Yamada}, S.,  \& {Sawada}, K.
  2008, \apj, 678, 1207

\bibitem[\protect\citeauthoryear{{Janka} et~al.}{{Janka}
  et~al.}{2007}]{JankaEtAl07}
{Janka}, H., {Langanke}, K., {Marek}, A., {Mart{\'{\i}}nez-Pinedo}, G.,  \&
  {M{\"u}ller}, B. 2007, \physrep, 442, 38

\bibitem[\protect\citeauthoryear{{Janka} \& {M\"{o}nchmeyer}}{{Janka} \&
  {M\"{o}nchmeyer}}{1989}]{JankaMoenchmeyer89}
{Janka}, H.,  \& {M\"{o}nchmeyer}, R. 1989, \aap, 209, L5

\bibitem[\protect\citeauthoryear{{Janka} \& {Mueller}}{{Janka} \&
  {Mueller}}{1996}]{JankaMuller96}
{Janka}, H.,  \& {Mueller}, E. 1996, \aap, 306, 167

\bibitem[\protect\citeauthoryear{{Kitaura}, {Janka}, \&
  {Hillebrandt}}{{Kitaura} et~al.}{2006}]{KitauraEtAl06}
{Kitaura}, F.~S., {Janka}, H.,  \& {Hillebrandt}, W. 2006, \aap, 450, 345

\bibitem[\protect\citeauthoryear{{Kotake}, {Yamada}, \& {Sato}}{{Kotake}
  et~al.}{2003}]{KotakeEtAl03}
{Kotake}, K., {Yamada}, S.,  \& {Sato}, K. 2003, \apj, 595, 304

\bibitem[\protect\citeauthoryear{{Kowarik} et~al.}{{Kowarik}
  et~al.}{2009}]{KowarikEtAl09}
{Kowarik}, T., {Griesel}, T., {Pi{\'e}gsa}, A.,  \& {for the Icecube
  Collaboration}. 2009, ArXiv e-prints, 0908.0441

\bibitem[\protect\citeauthoryear{{Leahy}, {Elsner}, \& {Weisskopf}}{{Leahy}
  et~al.}{1983}]{LeahyEtAl83}
{Leahy}, D.~A., {Elsner}, R.~F.,  \& {Weisskopf}, M.~C. 1983, \apj, 272, 256

\bibitem[\protect\citeauthoryear{{Liebend{\"o}rfer} et~al.}{{Liebend{\"o}rfer}
  et~al.}{2001}]{LiebendorferEtAl01}
{Liebend{\"o}rfer}, M., {Mezzacappa}, A., {Thielemann}, F., {Messer}, O.~E.,
  {Hix}, W.~R.,  \& {Bruenn}, S.~W. 2001, \prd, 63, 103004

\bibitem[\protect\citeauthoryear{{Liebend{\"o}rfer} et~al.}{{Liebend{\"o}rfer}
  et~al.}{2005}]{LiebendorferEtAl05}
{Liebend{\"o}rfer}, M., {Rampp}, M., {Janka}, H.,  \& {Mezzacappa}, A. 2005,
  \apj, 620, 840

\bibitem[\protect\citeauthoryear{{Livne} et~al.}{{Livne}
  et~al.}{2004}]{LivneEtAl04}
{Livne}, E., {Burrows}, A., {Walder}, R., {Lichtenstadt}, I.,  \& {Thompson},
  T.~A. 2004, \apj, 609, 277

\bibitem[\protect\citeauthoryear{{Lund} et~al.}{{Lund}
  et~al.}{2010}]{LundEtAl10}
{Lund}, T., {Marek}, A., {Lunardini}, C., {Janka}, H.,  \& {Raffelt}, G. 2010,
  ArXiv e-prints, 1006.1889

\bibitem[\protect\citeauthoryear{{Maeder} \& {Meynet}}{{Maeder} \&
  {Meynet}}{2000}]{MaederMeynet00}
{Maeder}, A.,  \& {Meynet}, G. 2000, \araa, 38, 143

\bibitem[\protect\citeauthoryear{{Marek} \& {Janka}}{{Marek} \&
  {Janka}}{2009}]{MarekJanka09}
{Marek}, A.,  \& {Janka}, H. 2009, \apj, 694, 664

\bibitem[\protect\citeauthoryear{{Marek}, {Janka}, \& {M{\"u}ller}}{{Marek}
  et~al.}{2009}]{MarekEtAl09}
{Marek}, A., {Janka}, H.,  \& {M{\"u}ller}, E. 2009, \aap, 496, 475

\bibitem[\protect\citeauthoryear{{Murphy} \& {Burrows}}{{Murphy} \&
  {Burrows}}{2008}]{MurphyBurrows08}
{Murphy}, J.~W.,  \& {Burrows}, A. 2008, \apj, 688, 1159

\bibitem[\protect\citeauthoryear{{Murphy}, {Ott}, \& {Burrows}}{{Murphy}
  et~al.}{2009}]{MurphyEtAl09}
{Murphy}, J.~W., {Ott}, C.~D.,  \& {Burrows}, A. 2009, \apj, 707, 1173

\bibitem[\protect\citeauthoryear{{Nordhaus} et~al.}{{Nordhaus}
  et~al.}{2010}]{NordhausEtAl10}
{Nordhaus}, J., {Burrows}, A., {Almgren}, A.,  \& {Bell}, J. 2010, ArXiv
  e-prints, 1006.3792

\bibitem[\protect\citeauthoryear{{Ott}}{{Ott}}{2009}]{Ott09}
{Ott}, C.~D. 2009, Classical and Quantum Gravity, 26, 204015

\bibitem[\protect\citeauthoryear{{Ott} et~al.}{{Ott} et~al.}{2008}]{OttEtAl08}
{Ott}, C.~D., {Burrows}, A., {Dessart}, L.,  \& {Livne}, E. 2008, \apj, 685,
  1069

\bibitem[\protect\citeauthoryear{{Ott} et~al.}{{Ott} et~al.}{2006}]{OttEtAl06}
{Ott}, C.~D., {Burrows}, A., {Thompson}, T.~A., {Livne}, E.,  \& {Walder}, R.
  2006, \apjs, 164, 130

\bibitem[\protect\citeauthoryear{{Rampp} \& {Janka}}{{Rampp} \&
  {Janka}}{2000}]{RamppJanka00}
{Rampp}, M.,  \& {Janka}, H. 2000, \apjl, 539, L33

\bibitem[\protect\citeauthoryear{{Scheck} et~al.}{{Scheck}
  et~al.}{2008}]{ScheckEtAl08}
{Scheck}, L., {Janka}, H., {Foglizzo}, T.,  \& {Kifonidis}, K. 2008, \aap, 477,
  931

\bibitem[\protect\citeauthoryear{{Shen} et~al.}{{Shen}
  et~al.}{1998a}]{ShenEtAl98a}
{Shen}, H., {Toki}, H., {Oyamatsu}, K.,  \& {Sumiyoshi}, K. 1998a, Nuclear
  Physics A, 637, 435

\bibitem[\protect\citeauthoryear{{Shen} et~al.}{{Shen}
  et~al.}{1998b}]{ShenEtAl98b}
{Shen}, H., {Toki}, H., {Oyamatsu}, K.,  \& {Sumiyoshi}, K. 1998b, Progress of
  Theoretical Physics, 100, 1013

\bibitem[\protect\citeauthoryear{{Shlomo}, {Kolomietz}, \& {Col{\`o}}}{{Shlomo}
  et~al.}{2006}]{ShlomoEtAl06}
{Shlomo}, S., {Kolomietz}, V.~M.,  \& {Col{\`o}}, G. 2006, European Physical
  Journal A, 30, 23

\bibitem[\protect\citeauthoryear{{Thompson}, {Burrows}, \& {Pinto}}{{Thompson}
  et~al.}{2003}]{ThompsonEtAl03}
{Thompson}, T.~A., {Burrows}, A.,  \& {Pinto}, P.~A. 2003, \apj, 592, 434

\bibitem[\protect\citeauthoryear{{Walder} et~al.}{{Walder}
  et~al.}{2005}]{WalderEtAl05}
{Walder}, R., {Burrows}, A., {Ott}, C.~D., {Livne}, E., {Lichtenstadt}, I.,  \&
  {Jarrah}, M. 2005, \apj, 626, 317

\bibitem[\protect\citeauthoryear{{Woosley}}{{Woosley}}{1993}]{Woosley93}
{Woosley}, S.~E. 1993, \apj, 405, 273

\bibitem[\protect\citeauthoryear{{Woosley}, {Heger}, \& {Weaver}}{{Woosley}
  et~al.}{2002}]{WoosleyEtAl02}
{Woosley}, S.~E., {Heger}, A.,  \& {Weaver}, T.~A. 2002, Rev. Mod. Phys., 74,
  1015

\bibitem[\protect\citeauthoryear{{Yakunin} et~al.}{{Yakunin}
  et~al.}{2010}]{YakuninEtAl10}
{Yakunin}, K.~N., et~al. 2010, ArXiv e-prints, 1005.0779

\bibitem[\protect\citeauthoryear{{Yamasaki} \& {Foglizzo}}{{Yamasaki} \&
  {Foglizzo}}{2008}]{YamasakiFoglizzo08}
{Yamasaki}, T.,  \& {Foglizzo}, T. 2008, \apj, 679, 607

\end{thebibliography}

\end{document}